\documentclass[twocolumn]{aastex631}
\usepackage{times,amsmath}
\usepackage[T1]{fontenc}

\hypersetup{pdfauthor={Ewan O'Sullivan},
            pdftitle={A hot-core group-dominant elliptical galaxy NGC 777},
            pdfkeywords={active galactic nuclei; circumgalactic medium; cooling flows; elliptical galaxies; galaxy groups; intracluster medium},
            bookmarksnumbered=true}
\pdfoutput=1

\newcommand{\arcm}{\hbox{$^\prime$}}

\newcommand{\degree}{\hbox{$^\circ$}}

\newcommand{\chandra}{\emph{Chandra}}
\newcommand{\xmm}{\emph{XMM-Newton}}
\newcommand{\xmms}{\emph{XMM}}

\newcommand{\arcs}{\mbox{\arcm\arcm}}

\newcommand{\Zsol}{\ensuremath{\mathrm{~Z_{\odot}}}}

\newcommand{\Msol}{\ensuremath{\mathrm{~M_{\odot}}}}
\newcommand{\Msolpyr}{\ensuremath{\mathrm{~M_{\odot}~yr^{-1}}}}

\newcommand{\NH}{\ensuremath{N_{\mathrm{H}}}}

\newcommand{\s}{\ensuremath{\mbox{~s}}}
\newcommand{\ps}{\ensuremath{\s^{-1}}}
\newcommand{\cm}{\ensuremath{\mbox{~cm}}}
\newcommand{\pcmsq}{\ensuremath{\cm^{-2}}}

\newcommand{\kev}{\ensuremath{\mbox{~keV}}}
\newcommand{\kevcmsq}{\ensuremath{\kev\cm^{2}}}
\newcommand{\km}{\ensuremath{\mbox{~km}}}
\newcommand{\Mpc}{\ensuremath{\mbox{~Mpc}}}
\newcommand{\pMpc}{\ensuremath{\Mpc^{-1}}}
\newcommand{\kmpspMpc}{\ensuremath{\km \ps \pMpc\,}}
\newcommand{\erg}{\ensuremath{\mbox{~erg}}}
\newcommand{\ergps}{\ensuremath{\erg \ps}}
\newcommand{\ergpspcmsq}{\ensuremath{\erg \ps \pcmsq}}
\newcommand{\kmps}{\ensuremath{\km \ps}}

\newcommand{\Dtf}{\ensuremath{D_{\mathrm{25}}}}

\newcommand{\gtsim}{\,\rlap{\raise 0.5ex\hbox{$>$}}{\lower 1.0ex\hbox{$\sim$}}\,}

\newcommand{\Ha}{\ensuremath{\mathrm{H\alpha}}}

\received{---}
\revised{---}
\accepted{---}
\submitjournal{ApJ}

\shorttitle{A hot core in NGC~777}
\shortauthors{O'Sullivan et al.}

\begin{document}

\title{
A hot core in the group-dominant elliptical galaxy NGC~777}

\correspondingauthor{Ewan O'Sullivan}
\email{eosullivan@cfa.harvard.edu}

\author[0000-0002-5671-6900]{Ewan O'Sullivan}
\affiliation{Center for Astrophysics $|$ Harvard \& Smithsonian, 60 Garden
  Street, Cambridge, MA 02138, USA}

\author[0000-0001-7509-2972]{Kamlesh Rajpurohit}
\affiliation{Center for Astrophysics $|$ Harvard \& Smithsonian, 60 Garden
  Street, Cambridge, MA 02138, USA}

\author[0000-0002-4962-0740]{Gerrit Schellenberger}
\affiliation{Center for Astrophysics $|$ Harvard \& Smithsonian, 60 Garden
  Street, Cambridge, MA 02138, USA}

\author[0009-0007-0318-2814]{Jan Vrtilek}
\affiliation{Center for Astrophysics $|$ Harvard \& Smithsonian, 60 Garden
  Street, Cambridge, MA 02138, USA}

\author{Laurence P. David}
\affiliation{Center for Astrophysics $|$ Harvard \& Smithsonian, 60 Garden
  Street, Cambridge, MA 02138, USA}

\author[0000-0003-1746-9529]{Arif Babul}
\affiliation{Department of Physics and Astronomy, University of Victoria, Victoria, BC, V8P1A1, Canada}
\affiliation{Infosys Visiting Chair Professor, Department of Physics, Indian Institute of Science, Bangalore 560012, India}

\author[0000-0001-6638-4324]{Valeria Olivares}
\affiliation{Astrophysics Branch, NASA-Ames Research Center, MS 245-6, Moffett Field, CA 94035, USA}

\author[0000-0001-5338-4472]{Francesco Ubertosi}
\affiliation{Dipartimento di Fisica e Astronomia, Universit\'{a} di Bologna, via Gobetti 93/2, I-40129 Bologna, Italy}
\affiliation{INAF, Osservatorio di astrofisica e Scienza dello Spazio, via P. Gobetti 93/3, I-40129 Bologna, Italy}

\author[0000-0002-3104-6154]{Konstantinos Kolokythas}
\affiliation{Centre for Radio Astronomy Techniques and Technologies, Department of Physics and Electronics, Rhodes University, P.O. Box 94, Makhanda 6140, South Africa}
\affiliation{South African Radio Astronomy Observatory, 2 Fir Street, Observatory 7925, South Africa}

\author[0000-0003-3165-9804]{Iurii Babyk}
\affiliation{Center for Astrophysics $|$ Harvard \& Smithsonian, 60 Garden
  Street, Cambridge, MA 02138, USA}

\author[0000-0002-3937-7126]{Ilani Loubser}
\affiliation{Centre for Space Research, North-West University, Potchefstroom 2520, South Africa}

\begin{abstract}
NGC~777 provides an example of a phenomenon observed in some group-central ellipticals, in which the temperature profile shows a central peak, despite the short central cooling time of the intra-group medium. We use deep \chandra\ X-ray observations of the galaxy, supported by  uGMRT 400~MHz radio imaging, to investigate the origin of this hot core. We confirm the centrally-peaked temperature profile and find that entropy and cooling time both monotonically decline to low values (2.62$^{+0.19}_{-0.18}$\kevcmsq\ and 71.3$^{+12.8}_{-13.1}$~Myr) in the central $\sim$700~pc. Faint diffuse radio emission surrounds the nuclear point source, with no clear jets or lobes but extending to $\sim$10~kpc on a northwest-southeast axis. This alignment and extent agree well with a previously identified filamentary \Ha +[N\textsc{ii}] nebula. While cavities are not firmly detected, we see X-ray surface brightness decrements on the same axis at 10-20~kpc radius which are consistent with the intra-group medium having been pushed aside by expanding radio lobes. Any such outburst must have occurred long enough ago for lobe emission to have faded below detectability. Cavities on this scale would be capable of balancing radiative cooling for at least $\sim$240~Myr. We consider possible causes of the centrally peaked temperature profile, including gravitational heating of gas as the halo relaxes after a period of AGN jet activity, and heating by particles leaking from the remnant relativistic plasma of the old radio jets.
\end{abstract}

\keywords{
active galactic nuclei --- circumgalactic medium --- cooling flows --- elliptical galaxies --- galaxy groups --- intracluster medium
}

\section{Introduction}
\label{sec:intro}
Around the turn of the millennium, high spatial and spectral resolution X-ray observations revealed the impact of cluster- and group-central radio galaxies on the hot halos in which they are embedded \citep[e.g.,][]{Bohringeretal93,Rizzaetal00,Fabianetal00,McNamaraetal00,Vrtileketal00,FinoguenovJones01,Heinzetal02}. At the same time, theoretical models began to demonstrate the importance of active galactic nuclei (AGNs) as a source of heating for these systems \citep[e.g.,][]{Babuletal02,McCarthyetal04}.
It is now widely accepted that radiative cooling of the intra-group and intra-cluster medium (IGrM and ICM) is balanced by the heating effects of radio jets launched by the central dominant galaxies of these systems \citep[e.g.,][]{Fabian12,McNamaraNulsen12,Gittietal12,Eckertetal21}. These jets are known to inflate lobes and drive shocks into the surrounding medium, causing heating directly or via the injection of sound waves and turbulence. In galaxy clusters, active radio jets and the products of ICM cooling (ionized or molecular gas, star formation) are most commonly seen in systems defined as strong cool cores, with low central entropies \citep[$\lesssim$30\kevcmsq, e.g.,][]{Voitetal08,Cavagnoloetal08}, central cooling times $<$1~Gyr, and central temperature declines \citep{Hudsonetal10}.

In galaxy groups (and individual ellipticals) the situation is made more complex by the increased efficiency of radiative cooling via line emission at the characteristic ($\sim$1~keV) temperature range of the IGrM \citep{OSullivanetal17}. Essentially all galaxy groups are found to have short central cooling times, in most cases $<$1~Gyr, regardless of their dynamical state or temperature profile. Indeed, there are observed cases of violently merging groups in which not only are the cool cores preserved, but there is evidence of cooling and ongoing jet activity in the central galaxy \citep[e.g.,][]{OSullivanetal19,Schellenbergeretal23}. The short cooling times in groups have obvious implications for the cycle of cooling and feedback, suggesting that the active galactic nuclei of their dominant galaxies will launch jets more often than their counterparts in clusters. However, the shallow potential wells of groups mean that AGN heating may have a greater impact on the intra-group medium than on cluster halos  (\citealp{Baloghetal99}; \citealp{Babuletal02}; see \citealp{Oppenheimeretal21} for a review).

Despite their short cooling times, observations of groups and ellipticals show a wide variety of temperature profiles. A profile similar to that of cool-core clusters (a low central temperature, peak at moderate radii and slow decline to larger radius) is common, but there are also systems where a central temperature increase is seen in the center of the cool core, or even where no cool core is seen, only a centrally-peaked temperature profile declining with radius. In a sample of 60 early-type galaxies in a range of environments, \citet{Kimetal20} find $\sim$35 per cent have centrally peaked temperature profiles (8 with monotonic negative gradients and 13 ``hybrid-dip'' profiles with negative inner gradients but positive outer gradients). \citet{OSullivanetal17} find central temperature peaks in 3 of 13 ($\sim$25 per cent) group-central ellipticals for which deprojected profiles were available.

Several explanations for such hot cores have been suggested. For the smallest, sub-kiloparsec scale temperature peaks, heating by confined jets or gravitational heating by the central supermassive black hole may be responsible \citep{Humphreyetal08,Pellegrinietal12,Paggietal14}. For sufficiently concentrated dark matter halos, gravitational heating is also a possibility on larger scales, with the gas in a global cooling flow gaining more energy than it can radiate away as it flows into the group core (\citealp{Khosroshahietal04,Humphreyetal06}; see also \citealp{Lewisetal00}). Working with a sample of ellipticals in a range of environments, \citet{Kimetal20} noted that central temperature peaks are associated with younger stellar ages, and suggested that star formation could be heating the core.

Finally, in some cases central temperature rises are associated with ongoing or recent AGN jet activity \citep[e.g.,][]{Machaceketal06,OSullivanetal07}. AGN feedback studies typically consider jets and cavities on scales of a few to a few hundred kiloparsecs \citep[e.g.,][]{Panagouliaetal14b} but it should be noted that even apparently radio quiet or quiescent systems can host small scale (sub-kiloparsec) jets \citep[e.g.,][]{Kharbetal23,Silpaetal23,Ubertosietal24}. It is rare to see direct evidence of heating by AGN, and the great majority of groups and clusters which host active central jet sources show central temperature declines even if only on small scales \citep[e.g.,][]{Sun09}. Even in systems where AGN-driven shocks are confirmed \citep[e.g.,][]{Randalletal11,Randalletal15,Bogdanetal14} or suspected \citep{OSullivanetal11c} a central decline in temperature is still usually seen, suggesting that the heating effect of these weak shocks is brief, with the energy they inject radiated away on timescales of $\sim$10~Myr. It should however be noted that in elliptical dominated, X-ray luminous groups, hot cores are most commonly observed in lower mass systems \citep{Buote17,OSullivanetal17,Kimetal20}, perhaps indicating that the impact of heating (by whatever mechanism) is only visible in their shallower potential wells, or that heating can only effectively overcome radiative cooling in  smaller halos.

Hot cores therefore raise several questions: how are they formed? If AGN are responsible how is the heating accomplished, given that in most groups and clusters AGN heating merely balances cooling without dramatically changing the central temperature profile? And how do hot cores affect our understanding of the cooling-AGN feedback loop? In this paper we describe deep X-ray and radio observations of the relaxed group-dominant galaxy NGC~777, which has a centrally peaked temperature profile, a filamentary \Ha +[N\textsc{ii}] nebula characteristic of IGrM cooling, but which showed only ambiguous signs of AGN feedback.

\subsection{NGC~777}

NGC~777 is a slow-rotating E1 elliptical galaxy \citep{RC3,Eneetal20,Loubseretal22} located at the center of the relatively low mass X-ray luminous galaxy group LGG~42 \citep[M$_{500}$$\sim$2.8$\times$10$^{13}$\Msol,][and references therein]{OSullivanetal17}. It hosts an AGN classed as Seyfert~2/LINER~2 from optical spectroscopy \citep{Hoetal97b}. Observations with the Very Large Array (VLA) at 1.4 and 4.8~GHz (in A and B configuration respectively) revealed a radio source coincident with the AGN and marginally extended on arcsecond scales, with an alignment roughly NW-SE \citep{HoUlvestad01}. Giant Metrewave Radio Telescope (GMRT) at 235 and 610~MHz \citep{Kolokythasetal18} found only an unresolved core, with fairly flat spectral indices\footnote{Where spectral index $\alpha$ is defined as $S_\nu\propto\nu^{\alpha}$, with $S_\nu$ the flux density at frequency $\nu$.} $\alpha^{\rm 610MHz}_{\rm 235MHz}$=-0.75$\pm$0.04 and $\alpha^{\rm 1400MHz}_{\rm 235MHz}$=-0.61$\pm$0.15, consistent with a small active jet or core emission. A VLA 15~GHz observation \citep{Saikiaetal18}, VLA Sky Survey 2-4~GHz imaging \citep{Lacyetal20} and eMERLIN 1.5~GHz imaging at $\sim$0.2\arcs\ resolution \citep{Baldietal18} also find only an unresolved core. However, \citet{Capettietal22}, using 120-168~MHz data from the LOFAR Two-metre Sky Survey \citep[LoTSS;][]{Shimwelletal17} discovered diffuse emission extending NW-SE on scales $\sim$50\arcs\ ($\sim$18~kpc). Given that this emission was not detected in the older GMRT data, it seems likely that its spectral index is steep and that it is therefore aged plasma no longer energised by AGN jets. However, artefacts associated with nearby bright sources overlap the diffuse emission, making an accurate determination of its extent difficult.

Spectral Energy Distribution (SED) fitting suggests that the galaxy has a low star formation rate, $\sim$0.2\Msolpyr\ \citep{Amblardetal14}, with estimates from infrared or ultraviolet fluxes even lower \citep{OSullivanetal15,Kolokythasetal22}. The luminosity weighted stellar age in the galaxy core \citep[5.4$\pm$2.1~Gyr,][]{Annibalietal07} is consistent with minimal recent star formation. Integral field spectroscopy from the Multi Unit Spectroscopic Explorer (MUSE) on the Very Large Telescope (VLT) revealed the presence of ionized gas in the galaxy core, with a clumpy ($\sim$10~kpc) filament extending NW and hints of smaller filaments also extending along a NW-SE alignment similar to that of the radio emission \citep{Olivaresetal22}. Given its low velocity relative to the galaxy, lack of velocity gradients or rotation, this ionized gas seems likely to be the product of radiative cooling from the surrounding hot intra-group medium (IGrM). However, no cooler material has yet been detected in the galaxy, with IRAM~30m observations placing an upper limit on the mass of molecular gas in the galaxy core of $<$2.08$\times$10$^8$\Msol\ \citep{OSullivanetal18b}.

Throughout the paper we adopt a flat cosmology with H$_0$=70\kmpspMpc, $\Omega_\Lambda$=0.7 and $\Omega_{\rm M}$=0.3. We adopt a redshift for NGC~777 of 0.016728 \citep{Denicoloetal05} and a distance of 73~Mpc to match \citet{OSullivanetal17}, giving an angular scale of 1\arcs=354~pc.

\section{Observations and Data Reduction}
\label{sec:obs}

\subsection{Chandra}
\label{sec:chandra}
An initial snapshot ($\sim$9.5~ks) ACIS-I observation of NGC~777 was performed by \chandra\ in Cycle~5, with deeper ($\sim$102~ks) ACIS-S observations following in Cycle~23. These observations are contained in the \chandra\ Data Collection (CDC) 212~\dataset[doi:10.25574/cdc.212]{https://doi.org/10.25574/cdc.212}. Table~\ref{tab:obs} provides some details of the observations, all of which were made using the VFAINT telemetry mode. A summary of the \chandra\ mission is provided in \citet{Weisskopfetal02}.

\begin{table}
\caption{\label{tab:obs}Summary of the \chandra\ observations}
\begin{center}
\begin{tabular}{lcccc}
\hline
\hline
ObsID & PI & Instrument & Observation & Cleaned \\
 & & & date & exposure (ks) \\
\hline
5001  & Murray     & ACIS-I & 2004-12-23 &  9.52 \\
25488 & O'Sullivan & ACIS-S & 2021-11-18 & 19.45 \\
25827 & O'Sullivan & ACIS-S & 2022-11-05 & 10.92 \\
25828 & O'Sullivan & ACIS-S & 2022-11-03 & 12.03 \\
26207 & O'Sullivan & ACIS-S & 2021-11-19 & 18.77 \\
26208 & O'Sullivan & ACIS-S & 2021-11-19 & 11.94 \\
27530 & O'Sullivan & ACIS-S & 2022-11-06 & 14.85 \\
27531 & O'Sullivan & ACIS-S & 2022-11-05 & 14.00 \\
\multicolumn{4}{r}{Total ACIS-S:} & 101.96 \\
\hline
\end{tabular}
\end{center}
\end{table}

We reduced and analysed the observations using \textsc{ciao} 4.15 \citep{Fruscioneetal06} and CALDB 4.10.2, following the approach laid out in \citet{OSullivanetal17} and the \chandra\ analysis threads\footnote{https://cxc.harvard.edu/ciao/threads/index.html}. Periods of high background were identified using the \textsc{lc\_clean} script and filtered out of the observations. The standard \chandra\ blank-sky event files were reprojected to match each dataset, renormalized to match their 9.5-12~keV count rate, and used to create background images and spectra.

All observations were reprojected to match the coordinates of the deepest exposure, ObsID~26207, based on the positions of point sources identified in the field of view. The \textsc{merge\_obs} task was used to create images, exposure maps and point spread function (PSF) maps in several energy bands, with the PSF extent set to enclose at least 90 per cent of the point source flux. Point sources were then identified from the combined 0.5-7~keV image using \textsc{wavdetect} and excluded from further analysis (with the exception of the falsely detected source coincident with the galaxy nucleus).

Image analysis was performed using combined images in the 0.5-2~keV band, which provides a higher signal-to-noise ratio than broader bands. Spectra were extracted from each observation separately using the \textsc{specextract} task. The cycle~23 spectra were then combined into single spectra for each extraction region using \textsc{combine\_spectra}. Comparisons of fits to the combined cycle~23 spectra with either simultaneous fits to the individual spectra from all ObsIDs, or fits to spectra combining the 2021 and 2022 observations separately showed little change in results, suggesting that combining the spectra is acceptable for our purposes. Spectra were binned to a minimum of 20 counts per bin and fitted using \textsc{Xspec} 12.13.0b \citep{Arnaud96}. We adopted the solar abundance ratios of \citet{GrevesseSauval98} and a Galactic hydrogen column of 4.59$\times$10$^{20}$\pcmsq, drawn from the HI4PI survey \citep{HI4PI16}.

\subsection{XMM-Newton}
\label{sec:XMM}
\xmm\ has observed NGC~777 twice (ObsIDs 0203610301 and 0304160301) with nominal exposures of $\sim$30~ks in each case. Unfortunately, both observations are badly affected by background flaring. \citet{OSullivanetal17}, working with data from the European Photon Imaging Camera (EPIC), found that less than 15~ks of the first observation was unaffected, and discarded the second observation entirely. We reduced data from the Reflection Grating Spectrometers (RGS) using the \xmms\ Science Analysis System (\textsc{sas} v20.0.0) following the approach of \citet{Fabianetal22}. Both datasets were processed using \textsc{rgsproc}, and lightcurves were extracted from events on CCD 9 with flag values of 8 or 16, outside an absolute cross-dispersion angle of 1.5$\times$10$^{-4}$, in bins of 200s. These show the same background flaring found in the EPIC data. We follow \citet{Fabianetal22} in selecting good time intervals to include any lightcurve bin in which the count rate is $<$0.3~ct.~s${-1}$, resulting in cleaned exposures of 14.0 and 14.9~ks respectively. However, we note that this includes periods of low-level flaring.

We extracted spectra using 95 per cent of the pulse height distribution and 95 per cent of the cross-dispersion PSF. As NGC~777 is relatively centrally concentrated, we used background spectra extracted from the observations. The spectra were combined across RGS cameras and datasets using \textsc{rgscombine} to create single spectra for each order. These were then fitted simultaneously, using the 7-22\AA\ ($\sim$0.56-1.77~keV) band.

\subsection{uGMRT}
\label{sec:uGMRT}
As reported in \citet{Kolokythasetal18} NGC~777 was observed by the Giant Metrewave Radio Telescope (GMRT) in dual-frequency 235/610~MHz during 2010 December. These observations reached rms noise levels of 0.40 and 0.15~mJy~bm$^{-1}$ at 235 and 610~MHz respectively, but showed only a point-like radio source coincident with the galaxy nucleus. The group was reobserved with the upgraded GMRT \citep[uGMRT,][]{Guptaetal17} in band 3 (300-500~MHz) on 2023 August 30, using the GMRT Wideband Backend (GWB; project code 44$\_$009). Narrow-band data using the old correlator were also collected. The 200~MHz bandwidth was divided into 4096 channels of width 48.8~kHz, and the total time on source was 6~hr. 3C~48 was used as the flux and phase calibrator. The realtime Radio Frequency Interference (RFI) filter system for the uGMRT \citep{Buchetal19} was used to mitigate any broadband RFI that might be present during the observation.

The data were processed using the Source Peeling and Atmospheric Modelling pipeline \citep[SPAM,][]{Intemaetal09}, following the approach described in \citet{Rajpurohitetal21}. The wideband data were split into six sub-bands, and the flux density of the calibrator set to the scale of \citet{ScaifeHeald12}. The data were then averaged, flagged, and the bandpass correction was applied. A global sky model based on the narrow-band data was then used as the starting point for self-calibration, and the resulting calibrated sub-band images were recombined to form the final full-band image.

The restoring beam of the full resolution band 3 image has a half-power beam width (HPBW) of 7.48$\times$5.41\arcs\ with position angle of 133.8\degree\ (anti-clockwise from West) and an rms noise level close to NGC~777 of 16.2~$\mu$Jy~bm$^{-1}$. However, a nearby radio source, B2~0156+31 (separation $\sim$6.7\arcm) is bright enough \citep[257~mJy at 1.4~GHz,][]{Condonetal98} to produce some low-level linear artefacts in the field around NGC~777, and we therefore used a 10\arcs\ resolution image created using uvtaper for most of our analysis. With the circular 10\arcs\ HPBW restoring beam, we find rms noise levels of 19.6~$\mu$Jy~bm$^{-1}$.

\section{Analysis and Results}
\label{sec:results}
\begin{figure}
\includegraphics[width=\columnwidth,bb=36 124 577 668]{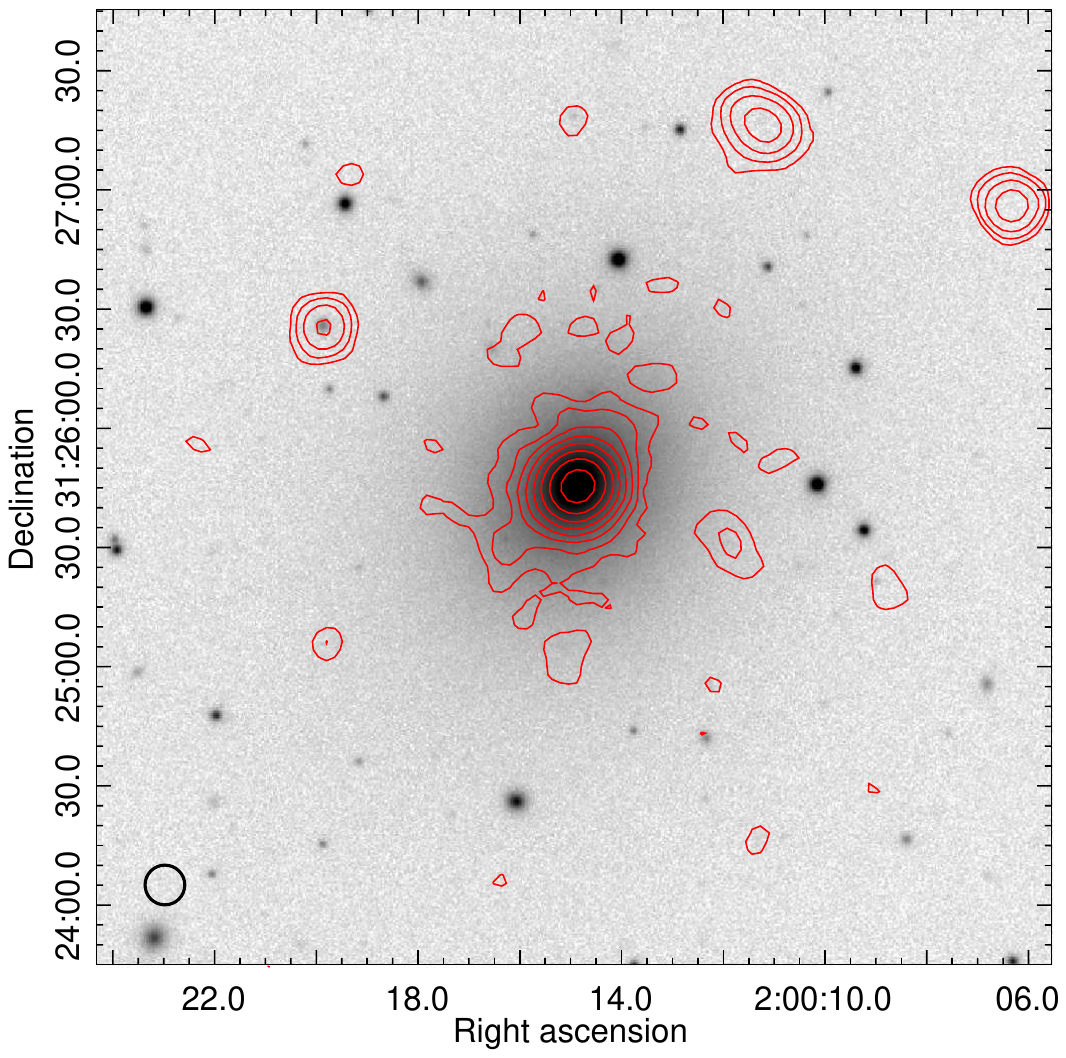}
\caption{\label{fig:opt_radio}SDSS $r$-band image of NGC~777 with uGMRT band 3 (410~MHz) contours overlaid. The radio image was created using Briggs weighting with robust parameter 0, and the 10\arcs\ restoring beam is indicated by the circle in the lower left corner. Contours begin at 78.4~$\mu$Jy~bm$^{-1}$ (4$\times$rms) and increase in steps of factor 2.}
\end{figure}

Figure~\ref{fig:opt_radio} shows an SDSS $r$-band image of NGC~777 overlaid with uGMRT band 3 (300-500~MHz) contours, created using a 10\arcs\ circular restoring beam. The contours show the previously observed radio core coincident with the nucleus of the galaxy, but also low-level diffuse emission extending to 20\arcs\ on all sides of the core, and to $\sim$28\arcs\ ($\sim$9.9~kpc) to the NW and SE. This is consistent with the full resolution uGMRT image, which shows extension on a scale of $\sim$25\arcs\ ($\sim$8.9~kpc) along a northwest-southeast (NW-SE) axis. We note that artefacts caused by the nearby bright source B2~0156+31 do not align with the NW-SE axis we see in the diffuse emission. Comparison with the LOFAR image of NGC~777 shows that the two datasets agree reasonably well on the extent and orientation of the diffuse emission, with the uGMRT image less affected by artefacts from B2~0156+31. We therefore conclude the extension along this axis is a real feature. We find the total 410~MHz flux density of NGC~777 to be 17.5$\pm$1.85~mJy, of which 3.2$\pm$0.4~mJy arises from the diffuse emission component. Uncertainties include a 10 per cent uncertainty on the absolute flux density scale of the uGMRT in band~3 \citep{ChandraKanekar17}, summed in quadrature with the rms noise over the area of the measurement.

\begin{figure}
\includegraphics[width=\columnwidth,bb=50 170 515 620]{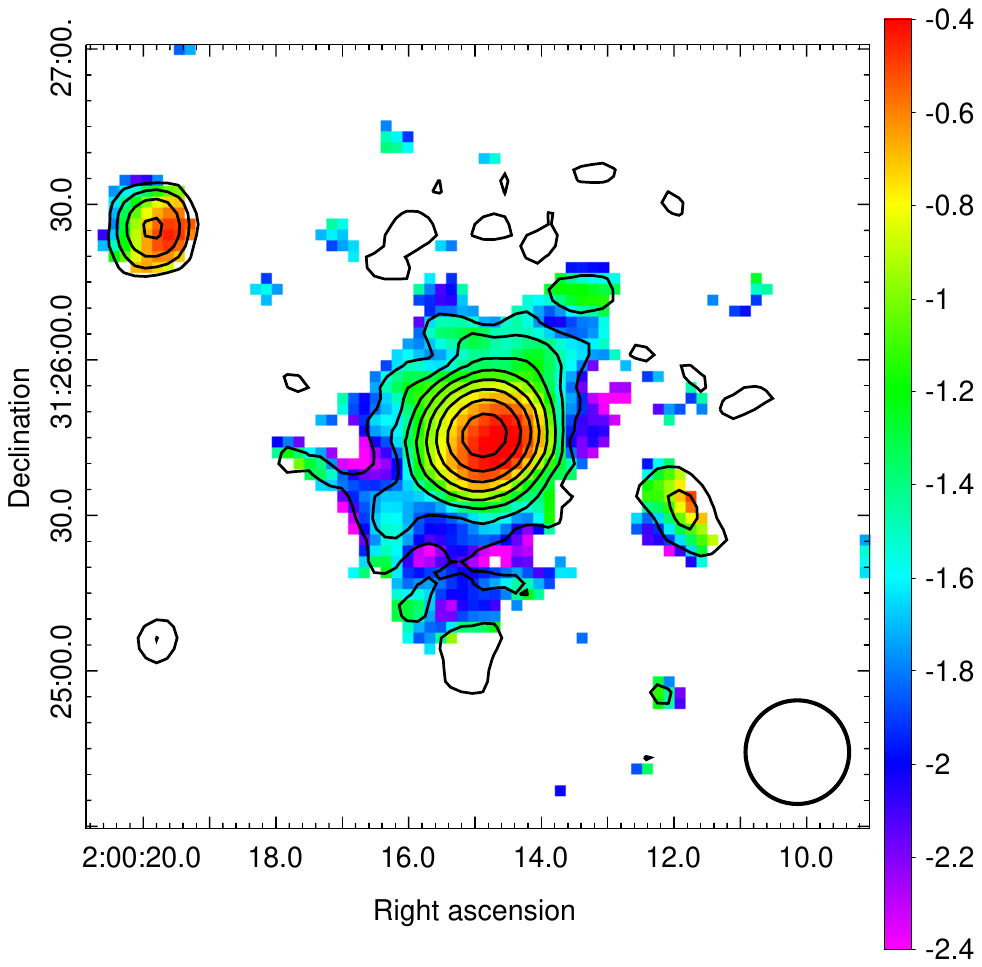}
\caption{\label{fig:spix}Map of radio spectral index between 410 and 144~MHz, based on uGMRT band 3 and LoTSS images clipped to remote regions of $<$3$\sigma$ significance. The 10\arcs\ restoring beam is indicated by the circle in the lower right corner. uGMRT band~3 contours are overlaid, starting at 78.4~$\mu$Jy~bm$^{-1}$ (4$\times$rms) and increasing in steps of factor 2.}
\end{figure}

To verify the flux density scale, we checked the spectra of compact sources in the field using our band~3 data, the LoTSS image, and the TGSS and NVSS survey images. The LOFAR 144~MHz flux density was found to be systematically high, and we therefore applied a flux correction factor of 0.8 to the LOFAR data. Figure~\ref{fig:spix} shows a map of the 144-410~MHz spectral index, after this correction. The unresolved radio core shows a flat index $\alpha^{410~MHz}_{140~MHz}$$\simeq$$-0.6$, consistent with ongoing low-level activity in the nucleus or unresolved small-scale jets. By contrast, the diffuse component has a steep index of $\sim$$-1.5$. This is generally consistent with old plasma left over from past jet activity.

\begin{figure*}
\includegraphics[width=\columnwidth,bb=55 150 560 650]{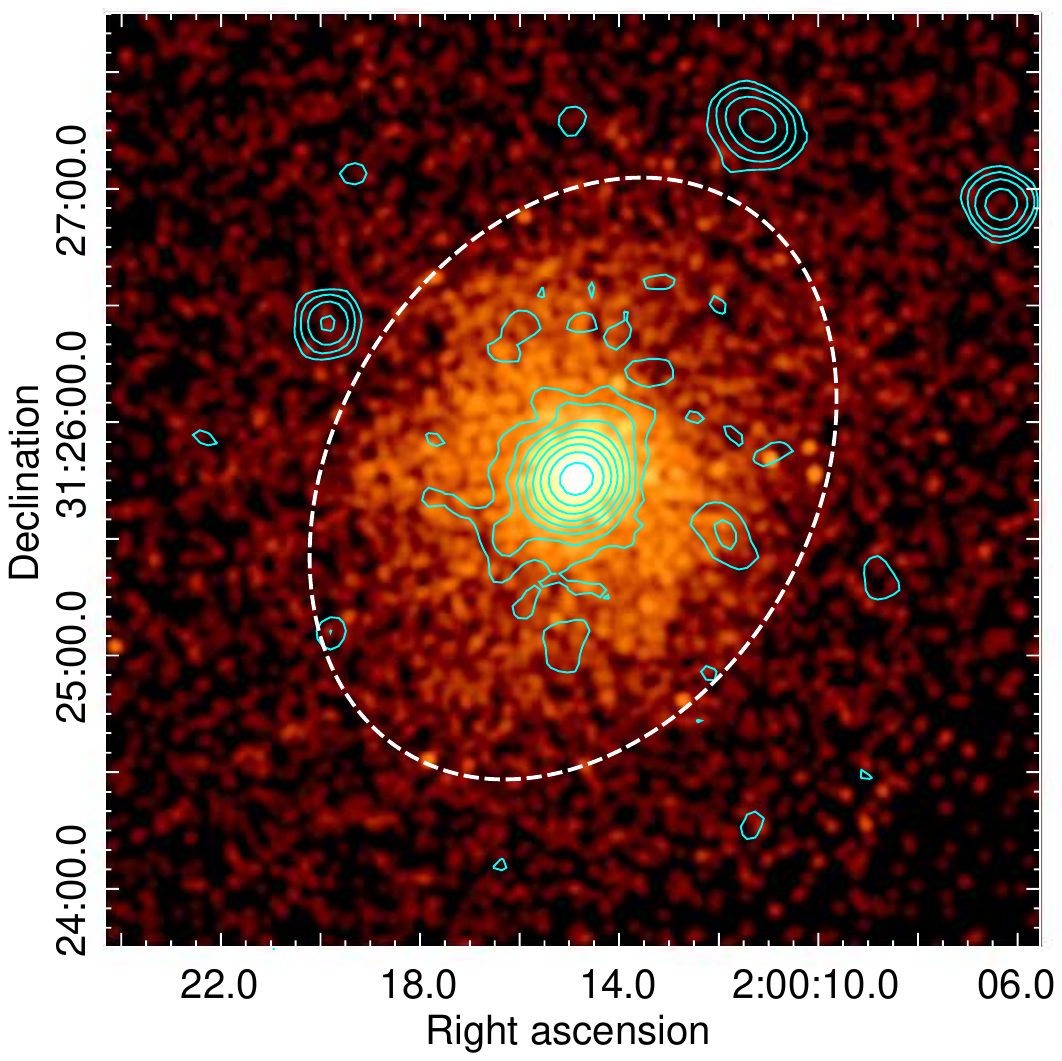}
\hspace{3mm}
\includegraphics[width=\columnwidth,bb=55 150 560 650]{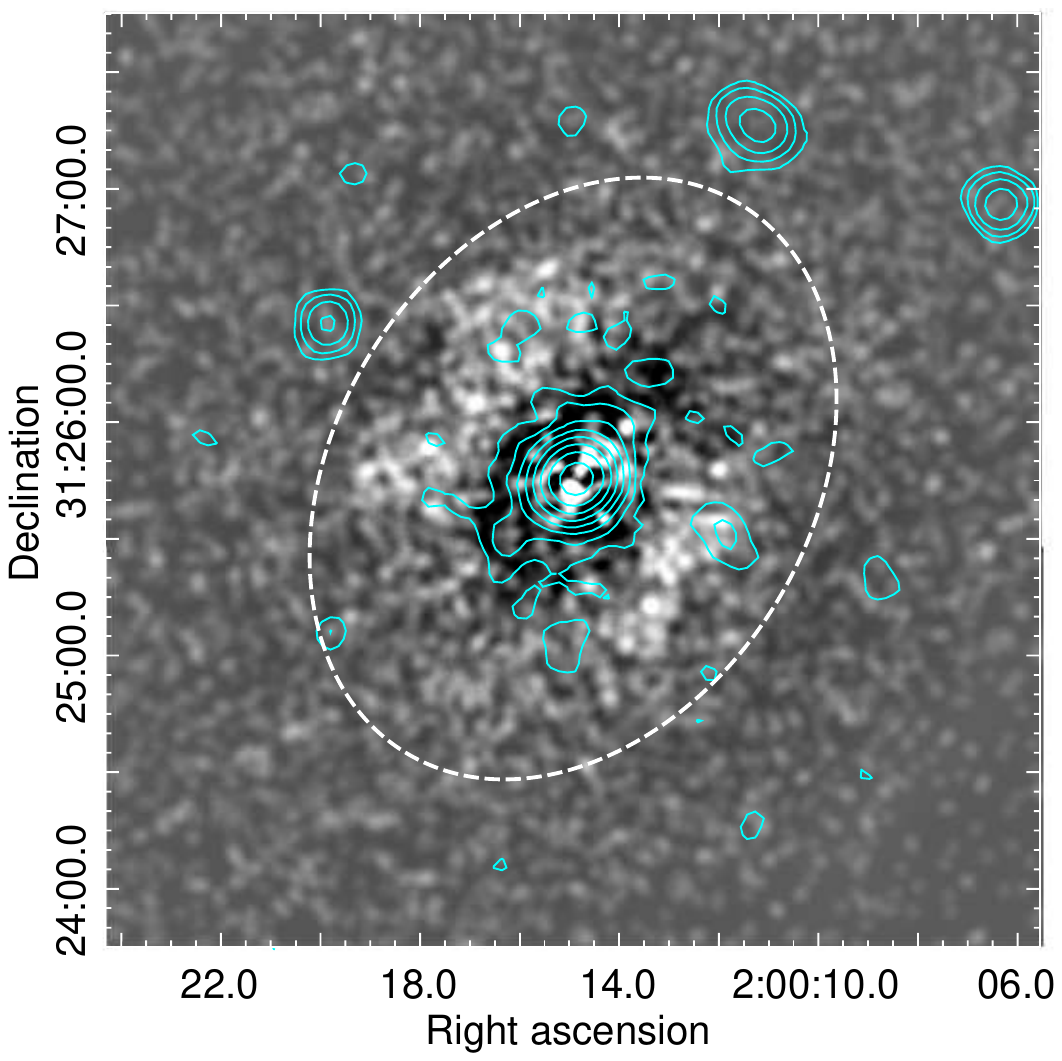}
\caption{\label{fig:Xrayims}\textit{Left}: \chandra\ 0.5-2~keV exposure-corrected image of NGC~777, smoothed with a 2.5\arcs\ radius Gaussian. uGMRT band~3 contours (as in Fig.~\ref{fig:opt_radio}) indicate the 400~MHz radio emission. The dashed ellipse shows the approximate \Dtf\ contour of the galaxy, with radii 82.6\arcs$\times$61.25\arcs\ (29.2$\times$21.7~kpc). \textit{Right}: Residual map after removal of point sources and subtraction of the best fitting 2D surface brightness model, showing the excess emission to the northeast and southwest of the core, and deficit along a northwest-southeast axis across the core.}
\end{figure*}

Figure~\ref{fig:Xrayims} (left panel) shows a lightly Gaussian-smoothed \chandra\ 0.5-2~keV image of NGC~777, combining all the available datasets. The X-ray emission peak closely matches the optical centroid of the galaxy and the core of the radio source. The diffuse X-ray emission shows significant structure within the stellar body of the galaxy (as indicated by the \Dtf\ ellipse) most notably brighter emission extending northeast-southwest (NE-SW) across the minor axis. On the NE side this shows a clear bifurcation toward its outer edge, with a concave edge between two more extended limbs of emission. The SW side also shows a somewhat concave outer edge, suggesting an overall X-shaped or ``butterfly winged'' structure. Along the major axis of the galaxy (NW-SE) the X-ray surface brightness falls more rapidly outside the central $\sim$25\arcs\ (8.9~kpc).

Previous studies using the cycle~5 \chandra\ snapshot have argued for the presence of cavities in the galaxy halo \citep{Cavagnoloetal10,Panagouliaetal14b,Olivaresetal22} aligned NE-SW, i.e., orthogonal to the alignment of the radio and \Ha +[N\textsc{ii}] structures. The locations of these suggested cavities match the concave outer edges of the X-shaped structure.

To investigate this further, we used \textsc{ciao} \textsc{sherpa} to model the diffuse emission, with the goal of subtracting the best fitting model to reveal any residual structure in the halo. We used a 0.5-2~keV image with point sources subtracted, folding the model through an exposure map. We initially used a single elliptical $\beta$-model to represent the group, with a flat component for the background. This produced an acceptable fit except in the galaxy nucleus, where the model leaves a small but extended, roughly circular residual. Attempting to fit this with a second $\beta$-model was unsuccessful, the model's outer slope becoming unphysically steep. However, a circular Gaussian provides a reasonable fit. The final parameters of the best fitting model are given in Table~\ref{tab:SBfit}.

\begin{table}
\caption{\label{tab:SBfit}Parameters of best-fitting surface brightness model. Position angles (P.A.) are measured anti-clockwise from west.}
\begin{center}
\begin{tabular}{lcc}
\hline
\hline
Component & Parameter & Value \\
\hline
$\beta$-model & R.A. & 02 00 14.85\\
& Dec. & +31 25 46.6\\
& r$_{\rm core}$ & 5.94$^{+0.42}_{-0.34}$\arcs\\
& $\beta$ & 0.49$\pm$0.13\\
& ellipticity & 0.076$\pm$0.013\\
& P.A. & 61.7$\pm$5.2\degree\\
Gaussian & R.A. & 02 00 14.84\\
& Dec. & +31 25 46.2\\
& FWHM & 3.14$^{+0.17}_{-0.15}$\arcs\\
\hline
\end{tabular}
\end{center}
\end{table}

Figure~\ref{fig:Xrayims} (right panel) shows the smoothed residual map after subtraction of this surface brightness model. This reveals a deficit along the (NW-SE) major axis of the galaxy, on scales out to at least 1\arcm\ ($\sim$21~kpc). The alignment of this deficit matches that of the radio emission. The outer parts of the X-shaped structure extending across the minor axis (the ``butterfly's wings'') are visible as two regions of excess emission, the northeastern part being somewhat more extensive. Both the radio and \Ha +[N\textsc{ii}] emission are largely confined within the central 25\arcs (8.9~kpc), falling within a region of low surface brightness residuals between the two ``wings'', and both follow the same NW-SE axis of this region.

\subsection{Spectral Mapping}
\label{sec:maps}

To map the 2-dimensional distribution of projected gas properties in the group core we adopt the approach described in \citet{OSullivanetal19}. We used a map pixel scale of 2.5\arcs $\times$2.5\arcs\ and circular spectral extraction region with radii chosen to include 500 net counts. Region size thus varies with surface brightness, from $\sim$2.5\arcs\ radii in the galaxy nucleus to $\sim$30\arcs\ radius at the edge of the maps. Spectra and responses were extracted from these regions and fitted with an absorbed APEC model, with hydrogen column fixed at the Galactic value. Abundance was poorly constrained, so was fixed at 0.75\Zsol, a representative value for the inner halo based on the radial profile fits described in Section~\ref{sec:profiles}. We note that varying this fixed abundance between 0.5 and 1\Zsol\ has no significant impact on the structures seen in the maps. Note that since the spectral extraction regions generally overlap, the map pixels are not independent and the resulting maps are analogous to adaptively smoothed images, with low surface brightness regions being more heavily smoothed. Comparison with our radial spectral fits shows no inconsistencies, indicating that the maps are reliable.

\begin{figure*}
\begin{center}
\includegraphics[height=8.8cm,bb=50 172 510 610]{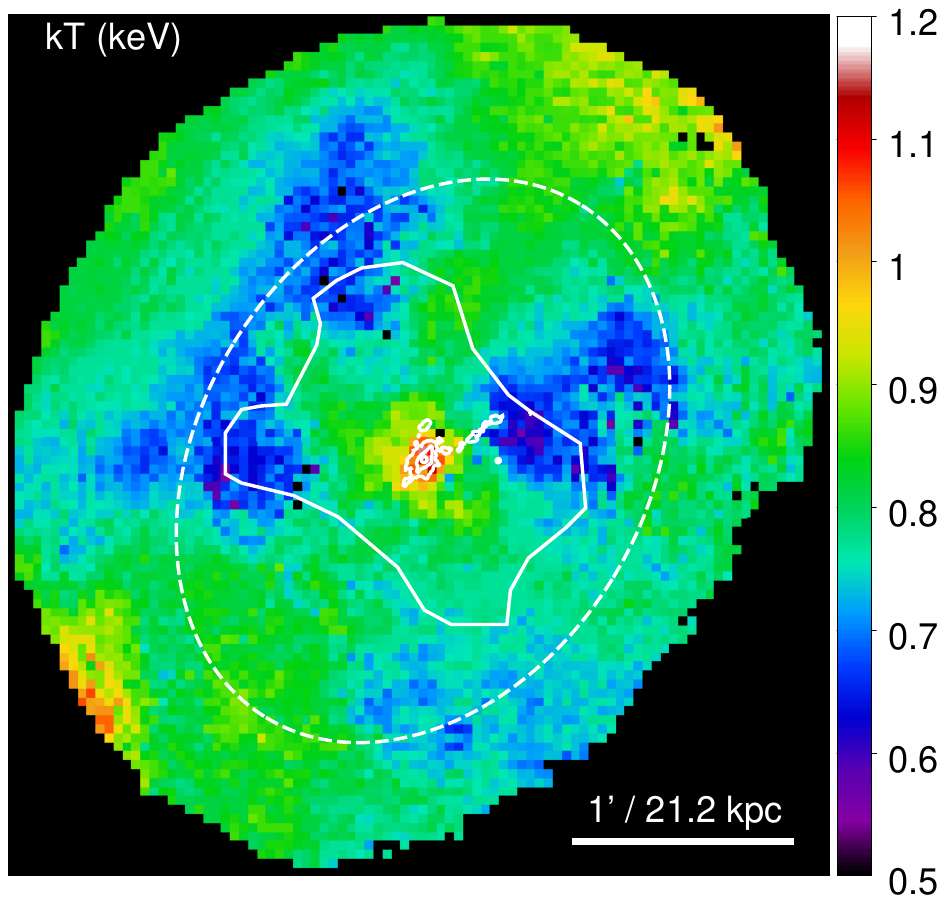}
\hspace{4mm}
\includegraphics[height=8.8cm,bb=50 115 562 665]{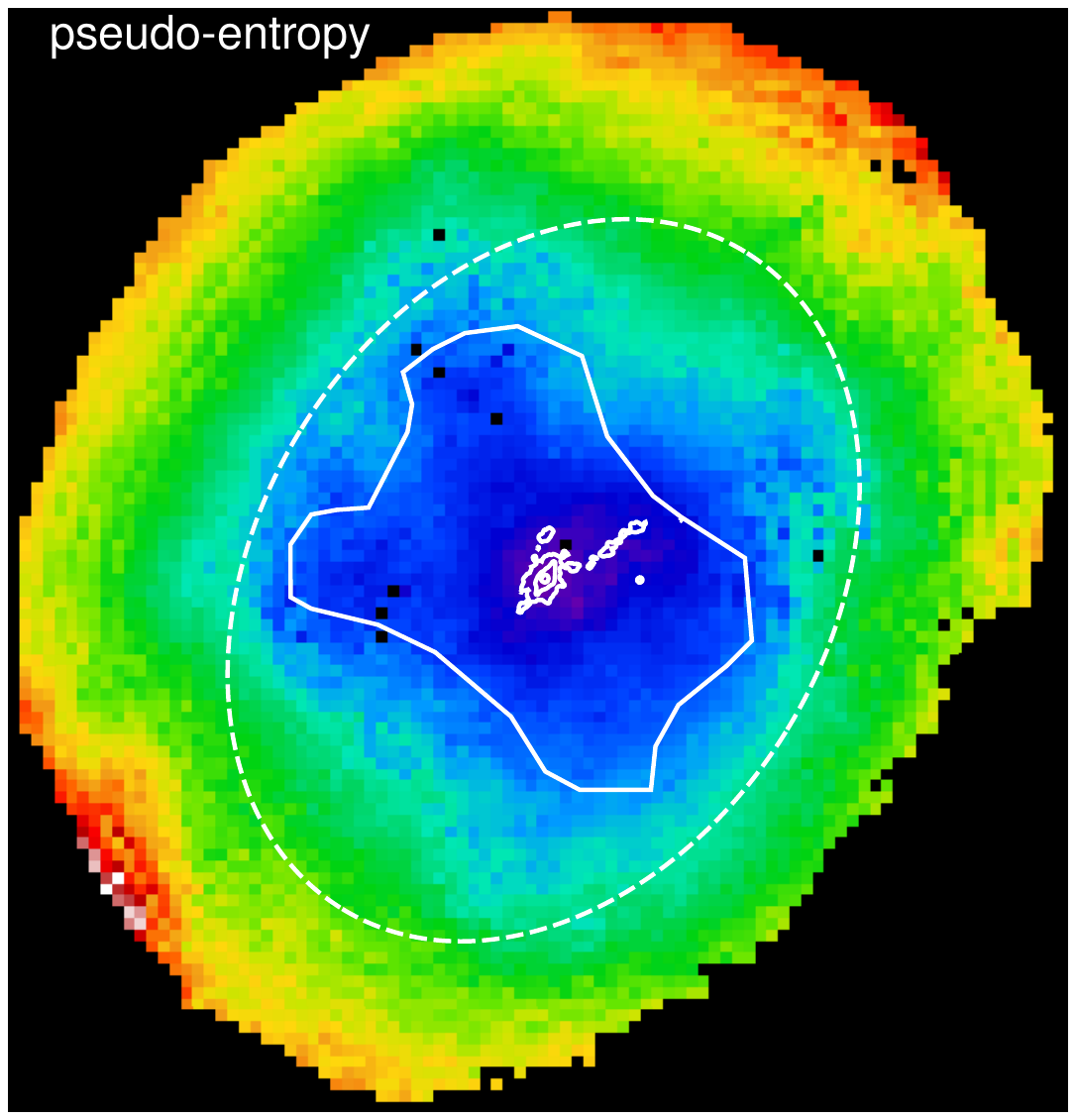}
\end{center}
\caption{\label{fig:kTmap}\chandra\ temperature and pseudo-entropy maps of NGC~777. The colour bar on the left panel indicates the temperature scale in keV. The pseudo-entropy map has arbitrary units, but the colour scale is linear and indicates an increase in entropy (from blue/purple to red) of a factor of $\sim$9. The optical extent of the galaxy is indicated by the dashed ellipse, which approximates the \Dtf\ contour. The location of the ionized gas filaments in the galaxy core are shown by MUSE [N\textsc{ii}] 6583~\AA\ contours, starting at 6$\sigma$ significance and increasing in steps of a factor of 6. The polygon indicates the extent of the butterfly wing structure seen in the surface brightness images.}
\end{figure*}

Figure~\ref{fig:kTmap} (left panel) shows the temperature map of the group core generated from the combined \chandra\ observations. Pixels in which the reduced $\chi^2$ of the spectral fit is greater than 1.5 or where the temperature uncertainty was greater than 20 per cent were excluded. Typical temperature uncertainties are $\sim$3-7 per cent in pixels warmer than $\sim$0.75~keV (green) rising to $\sim$16 per cent in the coolest pixels (0.6~keV, blue)


We also created pseudo-density, -pressure and -entropy maps, following the approach of \citet{Rossettietal07} in defining pseudo-density as the square root of the best-fitting normalization per unit area. Figure~\ref{fig:kTmap} (right panel) shows the pseudo-entropy map.

The temperature map shows clearly that the temperature structure in and around NGC~777 is not azimuthally symmetric. The central temperature peak is visible, but at larger radii we see moderate temperatures ($\sim$0.85~keV) extending NW and SE to warmer ($\sim$0.95~keV) regions at the edge of the map, while to the southwest and northeast the core is surrounded by cooler ($\sim$0.65~keV) regions, with temperature variations within them perhaps indicating clumps of cool gas. It is notable that the coolest temperatures ($\sim$0.6~keV) are seen NW of the core, and that the longest \Ha +[N\textsc{ii}] filament extends to the inner edge of this cool region. The pseudo-entropy map shows the lowest values in the core, with an overall increase with radius in all directions, but the cool clumps seen in the temperature map produce a degree of asymmetry, with a somewhat boxy overall entropy structure. The \Ha +[N\textsc{ii}] filaments are contained within the lowest entropy regions of the map. This, considered with their location in the galaxy centre, alignment with the radio and X-ray structures, low velocity relative to the galaxy, and lack of correlation with the stellar component, suggests that the ionized gas may have condensed out of the hot IGrM. Acquisition of the gas during a galaxy merger, or of gas stripped from satellite galaxies \citep[e.g.,][]{Saeedzadehetal23} appears less likely.

Comparing the temperature map and surface brightness images, it is clear that the cool regions of the map overlap the ``butterfly wing'' structures, but extend beyond them. The NW cool region in particular is located almost entirely outside the SW ``wing'', which is smaller than its NE counterpart. The path of higher temperatures extending SE from the core runs through the lower surface brightness region between the two ``wings'' on that side of the core, but the equivalent path to the NW overlaps the tip of the NE ``wing''. Overall, the spectral maps suggest that the ``butterfly wings'' largely consist of cooler, denser gas than their surroundings, but that there may be some degree of temperature structure along the line of sight that complicates the correlation. The smoothing effect of the adaptive mapping algorithm also needs to be borne in mind when interpreting the maps.

\subsection{Radial Spectral Profiles}
\label{sec:profiles}
\begin{figure*}
\includegraphics[width=0.49\textwidth,bb=30 220 565 760]{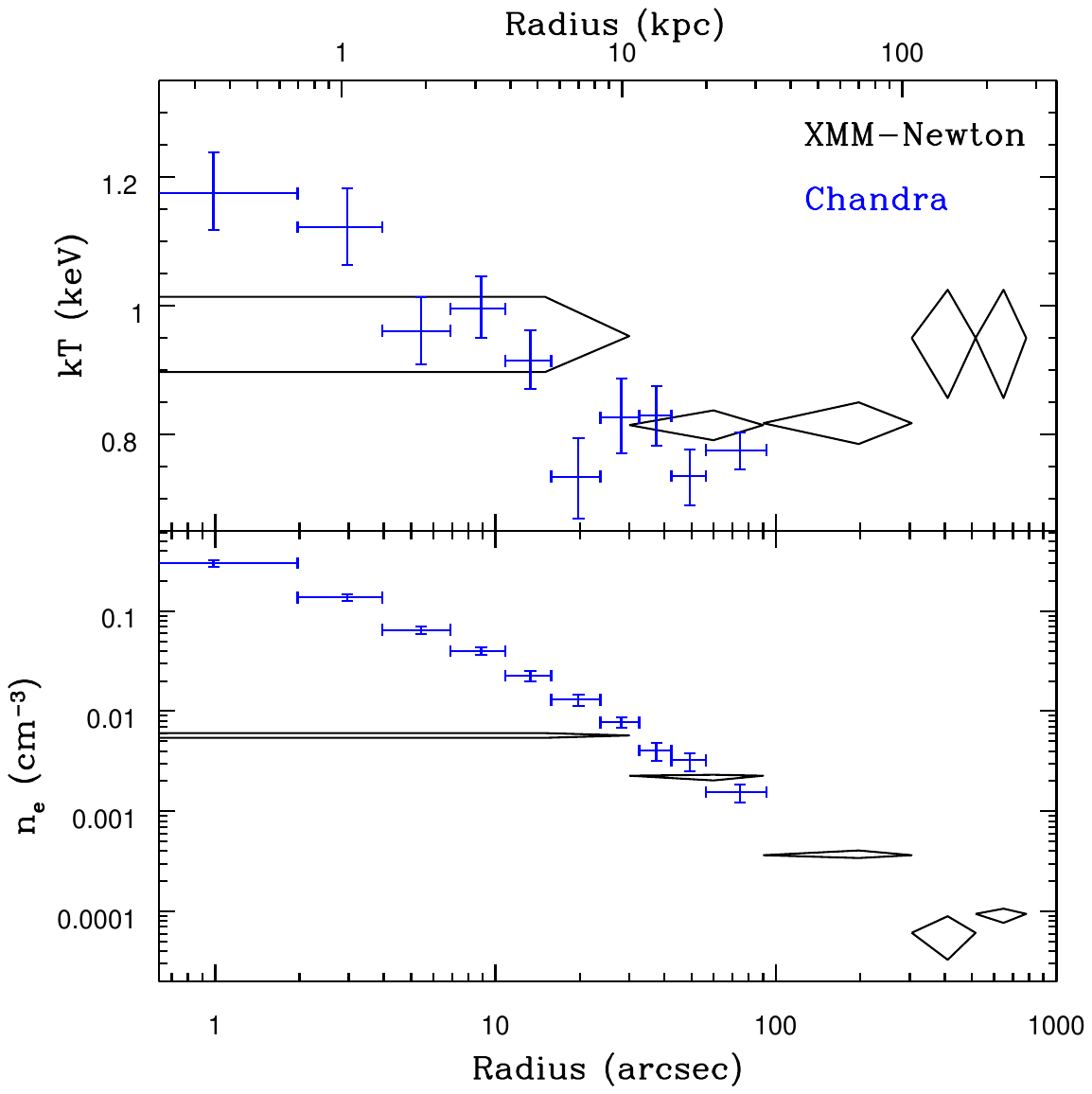}
\includegraphics[width=0.49\textwidth,bb=30 220 565 760]{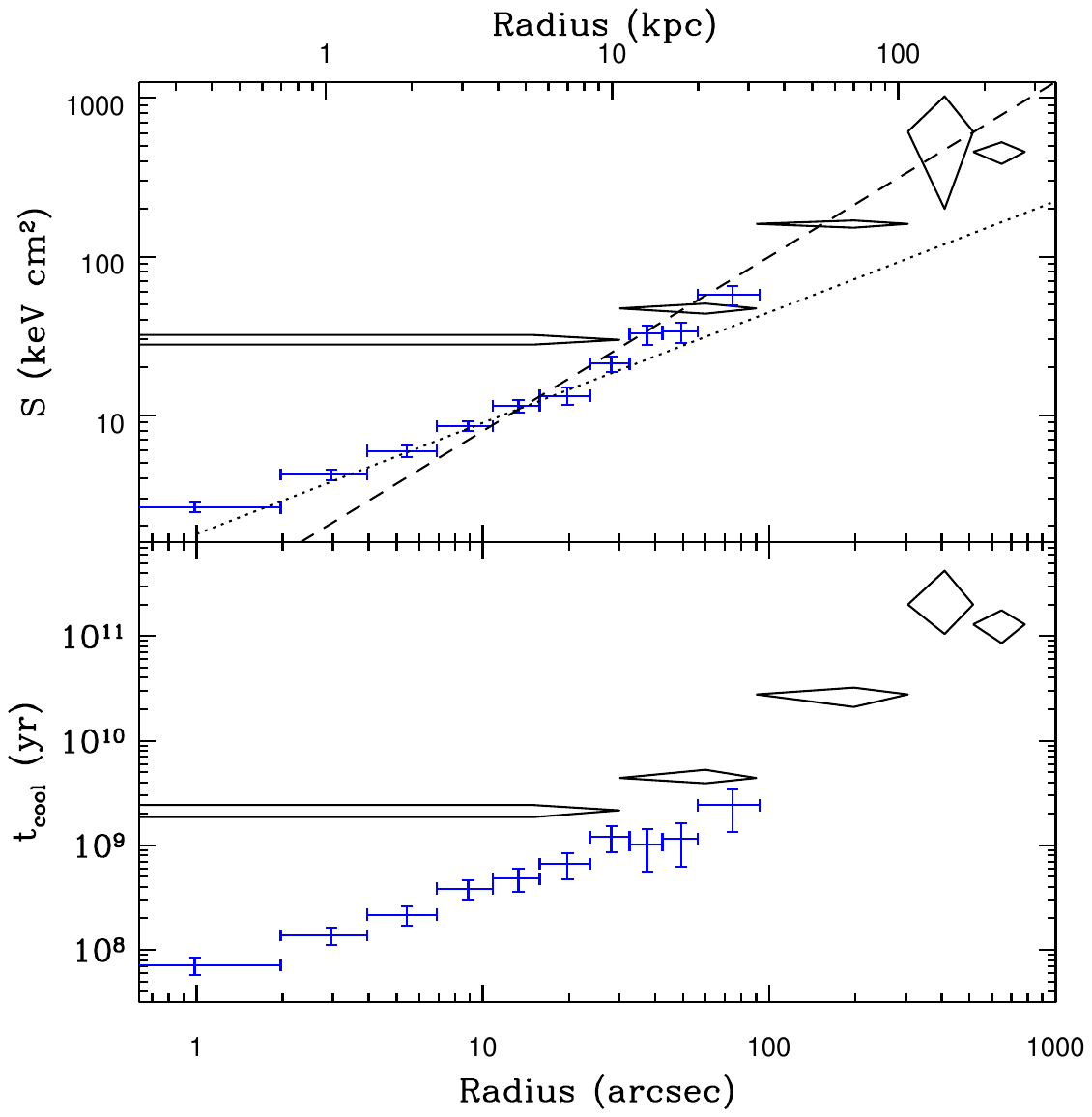}
\caption{\label{fig:deproj}Deprojected radial profiles of IGrM temperature, density, entropy and isobaric cooling time, with \chandra\ values marked by blue error bars and \xmms\ values \citep[from][]{OSullivanetal17} by black diamonds. The vertical extent of the symbols indicate 1$\sigma$ uncertainties. The dotted and dashed lines in the entropy panel show arbitrarily normalized gradients of R$^{0.7}$ and R$^{1.1}$ respectively.}
\end{figure*}

To determine the azimuthally averaged radial profiles of IGrM properties, we extracted spectra from annuli centered on the optical centroid and X-ray peak of the galaxy. We used circular annuli, since the ellipticity found by the surface brightness model is relatively small, and the brighter emission in the butterfly wing structures seems closer to circular; in any case we note that there are significant asymmetries in both brightness and temperature. Ten annuli were used, with widths chosen to achieve signal-to-noise ratios (S/N) of 35 for the spectra and a minumum width of 2\arcs\ (708~pc). We fitted the spectra with a deprojected absorbed thermal plasma model (\textsc{projct*phabs*apec}). The absorption was held at the Galactic value, and as the abundance was not well constrained in some individual annuli, we linked its value across groups of annuli. Pressure, entropy and isobaric cooling time were calculated from the fitted temperature electron density values \citep[as described in][]{OSullivanetal17}. The resulting temperature, density, entropy and cooling time profiles are shown in Figure~\ref{fig:deproj}.

To test for the presence of X-ray emission from the AGN, we also tried including a powerlaw component in the central bin, but this was found to be poorly constrained with a normalization consistent with zero. Since we also see no clear central point source in hard-band images, we conclude that the AGN is not X-ray luminous. To determine an upper limit on its flux we performed a projected fit to the spectrum in the central 2\arcs\ with abundance fixed at solar and powerlaw index fixed at $\Gamma$=1.7. The resulting 3$\sigma$ upper limit on AGN 0.5-7~keV flux is 1.31$\times$10$^{-14}$\ergpspcmsq, equivalent to a luminosity of L$_{0.5-7~keV}$$\leq$8.36$\times$10$^{39}$\ergps.


The deprojected temperature profile confirms the central peak seen in prior observations, with temperature continuing to rise down to scales below 1~kpc. Interestingly, both entropy and cooling time show smooth declining profiles, falling to 2.62$^{+0.19}_{-0.18}$\kevcmsq\ and 71.3$^{+12.8}_{-13.1}$~Myr in the central $\sim$700~kpc radius bin. The entropy at 10~kpc, K$_{10}$, which is often used to compare the state of different systems, is 32.61$^{+4.37}_{-5.01}$\kevcmsq.

Comparing with the \xmms\ profiles measured by \citet{OSullivanetal17}, the \chandra\ temperature profile is generally consistent with the \xmms\ profile and the entropy profiles overlap at $\sim$75\arcs\ radius. The \chandra\ cooling time profile falls a little below the \xmms\ profile, but at $\sim$75\arcs\ they are consistent within 2$\sigma$ uncertainties. The outer entropy profile traced by \xmms\ is consistent with an R$^{1.1}$ scaling, as expected for self-similarly scaling systems and the outer parts of groups and clusters where cooling and non-gravitational heating effects are minimal. In the core the profile is still expected to follow a power law, but the index will be dependent on the impact of heating and cooling, and the shape of the potential \citep{McCarthyetal05}. We find that in the central 10-20\arcs\ ($\sim$3.5-7.1~kpc) the \chandra\ entropy profile is well-described by an R$^{0.7}$ scaling, similar to the cores of many other galaxy clusters and groups \citep{Panagouliaetal14,Babyketal18}.

\subsection{RGS spectra}
The RGS data provide an alternate view of the X-ray emission from the core of the group, covering a band $\sim$1.7\arcm\ across centred on NGC~777. While the RGS spectra include emission from regions with different gas properties, their superior spectral resolution allows the emission lines of the Fe-L series to be separated from the continuum, providing additional constraints. \citet{Fabianetal22} used this capability to test the plausibility of a ``hidden cooling flow'' (HCF) model, in which the soft X-ray emission from the cooling gas is partially absorbed by the cold material that has condensed out of the ICM \citep{AllenFabian98}. The spectral resolution of RGS makes it possible to detect faint emission lines and thus to trace the cooling gas. This model has been applied to a number of clusters, groups and individual galaxies \citep{Fabianetal22,Fabianetal23a,Fabianetal23b} finding cooling rates of a few \Msolpyr\ in the group-dominant galaxies.


\begin{deluxetable*}{lcccccccccc}
\tablewidth{0pt}
\tablecaption{\label{tab:rgs}Model parameters for fits to the RGS spectra}
\tablehead{
\colhead{Model} & \colhead{kT$_1$} & \colhead{kT$_2$} & \colhead{Abund.} & \colhead{$z$} & \colhead{$\sigma_1$} & \colhead{$\sigma_2$} & \colhead{\NH} & \colhead{f$_{cov}$} & \colhead{$\dot M$} & \colhead{$\chi^2$/d.o.f} \\
 & \colhead{(keV)} & \colhead{(keV)} & \colhead{(Z$_\odot$)} & \colhead{(10$^{-2}$)} & \colhead{(10$^{-2}$ keV)} & \colhead{(10$^{-2}$ keV)} & \colhead{(10$^{22}$\pcmsq)} & & \colhead{(M$_\odot$~yr$^{-1}$)} & 
}
\startdata
\multicolumn{11}{l}{\textit{W statistic}}\\
2-T & 0.98$\pm$0.05 & 0.68$\pm$0.03 & 1.30$\pm$0.71 & 1.673$\pm$0.048 & $<$0.48 & 3.60$\pm$0.58 & - & - & - & 4231.5/3871 \\
HCF & 0.84$^{+0.02}_{-0.01}$ & - & 1\tablenotemark{a} & 1.650$^{+0.009}_{-0.029}$ & 1.34$^{+0.35}_{-0.50}$ & 4.74$^{+0.61}_{-1.20}$ & 3.00$^{+0.70}_{-0.22}$ & 0-1\tablenotemark{b} & 37.9$^{+1.9}_{-2.8}$ & 4248.7/3871 \\
\multicolumn{11}{l}{\textit{$\chi^2$ statistic}}\\
2-T & 1.32$^{+0.25}_{-0.14}$ & 0.76$\pm$0.03 & 1\tablenotemark{a} & 1.656$^{+0.058}_{-0.054}$ & = $\sigma_2$ & 2.95$\pm$0.45 & - & - & - & 312.3/272 \\
HCF & 0.85$\pm$0.02 & - & 1\tablenotemark{a} & 1.650$^{+0.068}_{-0.020}$ & = $\sigma_2$ & 3.16$^{+0.48}_{-0.46}$ & 3.47$^{+14.04}_{-1.39}$ & 0-1\tablenotemark{b} & 37.3$^{+1532.2}_{-18.4}$ & 308.7/271 \\
\enddata
\tablenotetext{a}{Abundance is poorly constrained and therefore fixed at solar.}
\tablenotetext{b}{Covering fraction is unconstrained.}
\end{deluxetable*}

\begin{figure}
\includegraphics[height=\columnwidth,bb=60 50 570 700,angle=-90]{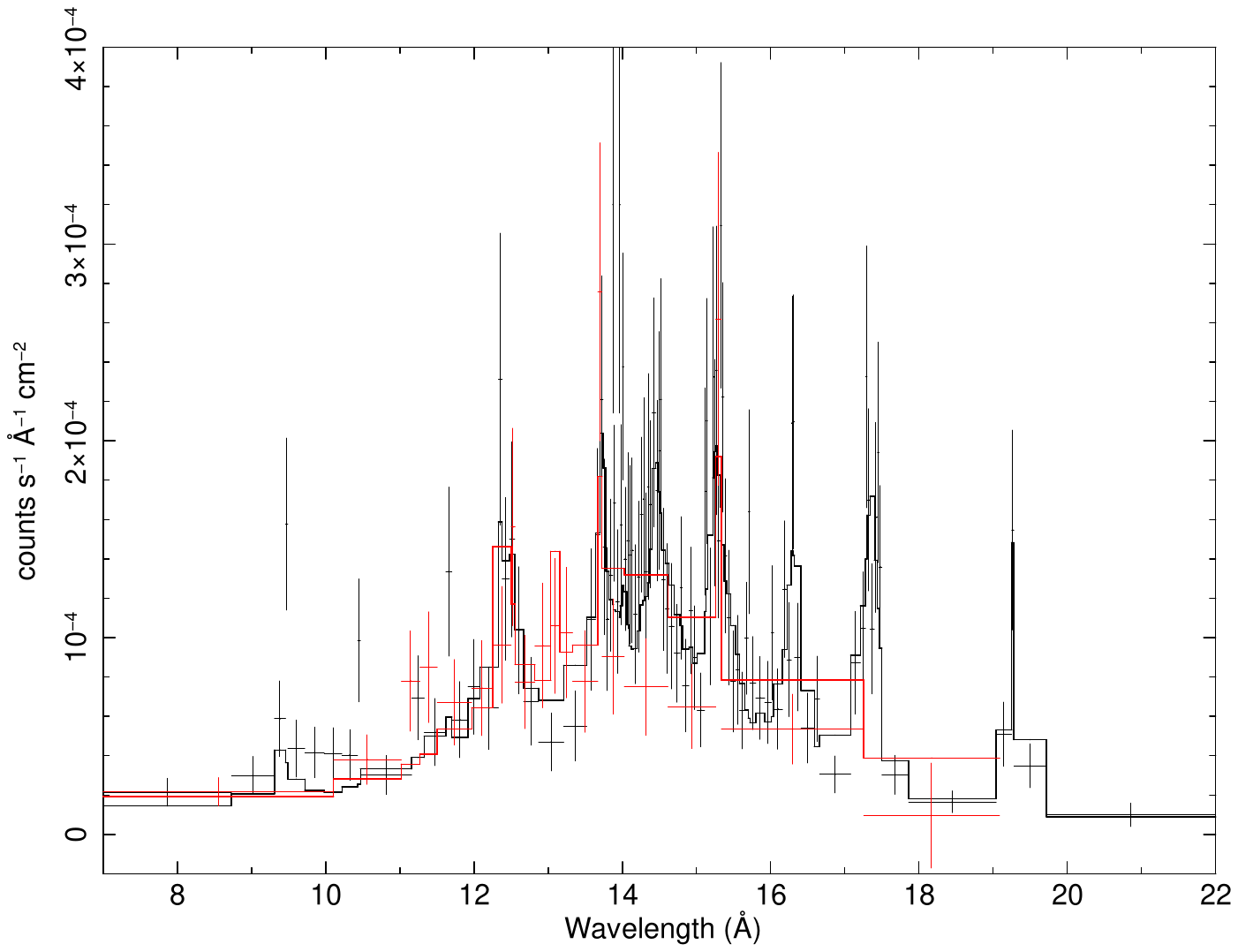}
\caption{\label{fig:rgs}\xmm\ RGS order 1 (black) and 2 (red) spectra in the 7-22\AA\ band, binned to 3$\sigma$ significance, with lines indicating the 2-temperature APEC model fit using the W statistic.}
\end{figure}

For NGC~777, the RGS spectra will include emission from regions with temperatures between $\sim$0.7 and $\sim$1.2~keV in the central $\sim$50\arcs\ of the group. We therefore fitted two models. The simpler model, representing a relatively stable IGrM with no cooling flow, is an absorbed 2-temperature APEC model, with each component smoothed to account for the blurring in the energy scale caused by the extended nature of the target, (\textsc{phabs*(gsmooth*apec+gsmooth*apec)}. Although our deprojection suggests a distribution of temperatures, we find that two temperatures is enough to fit the RGS spectra, and additional components provide no improvement. The second model is the HCF, as described in \citet{Fabianetal23a}, \textsc{tbabs*(gmooth*apec+ gsmooth(partcov*mlayerz)mkcflow)} where  \textsc{mlayerz} is a multiplicative model component defined as (1-\textsc{ztbabs})/(-ln(\textsc{ztbabs})) \citep{Fabianetal22}. In both cases the column of the initial absorption component was fixed at the Galactic value, and the abundance of the two components was linked. The degree of smoothing is defined at 6~keV with the index of the smoothing models fixed at 1, and \citet{Fabianetal22} note that smoothing the two components separately is necessary in the HCF model because the cooling flow is more centrally concentrated than the hotter phase of the ICM, and therefore requires a smaller smoothing kernel. In the HCF model, the upper temperature bound of the cooling flow component is linked to the temperature of the APEC component (kT$_1$) and the lower bound fixed at 0.1~keV. We found that the abundance of the HCF model was poorly constrained and rose to unphysical values, and we therefore fixed it at solar. An example of the spectra, fitted with the two-temperature model, is shown in Figure~\ref{fig:rgs}.

The relative faintness of NGC~777 and the short RGS exposures mean that we are faced with a trade-off between binning the spectra and retaining the best possible spectral resolution to support fitting of spectral lines. We therefore take two approaches, either binning the spectra to a minimum of 1 count per bin and fitting using the W-statistic, or binning to 20 counts per bin and using $\chi^2$ as the fit statistic. The best fitting parameter values and 1$\sigma$ uncertainties (or 3$\sigma$ upper limits) from our fits are shown in table~\ref{tab:rgs}. Both models provide fairly good fits to the data with comparable reduced $\chi^2$ values. In the 2-temperature APEC model, the two temperatures (and, when using the W-statistic, the abundance) are similar to the results found from the \chandra\ radial deprojection. For the HCF, the partial covering fraction (f$_{cov}$) is unconstrained even with abundance fixed, suggesting that the model can fit the spectra equally well with or without intrinsic absorption. In both models, when fitting with the W-statistic the smoothing kernel for the hotter component ($\sigma_1$) is smaller than that of the cooler component ($\sigma_2$) requires significant smoothing; in the 2-T model $\sigma_1$ is consistent with zero. This is the opposite of the behavior found by \citet{Fabianetal23a}, and suggests that the cooler component is more extended while the hotter component is compact. While this is consistent with the temperature profile, it argues against a cooling flow model, where cooling rate should peak strongly in the higher-density core. With the heavier binning of the $\chi^2$ fit, $\sigma_1$ is poorly constrained and consistent with $\sigma_2$ and we therefore link the two smoothing parameters. The bounds on the intrinsic absorption and cooling rate of the HCF model also extend to higher values. In all cases, the redshift is consistent with our adopted redshift for the system.

Given the dependence of the relative fit quality of the two models on the binning and fitting approach, it is difficult to rule out either with the current data. Both models reproduce all the visible emission lines comparably well, but testing shows that even a 1-temperature (0.79~keV, 0.65\Zsol) APEC model does reasonably well in this regard. The poor constraints on the HCF covering fraction and the apparent broader smoothing (and thus angular scale) of the cooler component when fitting with minimal binning are problematic for the HCF model, in that they suggest that the cooler gas may be more extended and that there are viable fitting solutions without significant intrinsic absorption. The 2-temperature model is better constrained, and provides an an acceptable fit, but is certainly an oversimplification of the true emission. We conclude that the RGS data alone do not provide a clear answer as to the state of the IGrM, and that higher quality high spectral resolution observations would be needed to provide better constraints.

\section{Discussion}
Combining the results of our radio and X-ray observations with previous findings, it appears that the AGN of NGC~777 may have had an outburst at some time in the past, but is now quiescent. The surrounding IGrM is therefore likely to be cooling and relaxing from the effects of this outburst.  Our radio observations confirm the presence of extended emission around the AGN, aligned along the same northwest-southeast axis we see in X-ray surface brightness. X-ray spectral mapping shows that the IGrM is not azimuthally symmetric, with cooler, denser gas in wings to the northeast and southwest, and lower surface-brightness, hotter material along the same axis as the radio emission. These correlations suggest a past AGN outburst which pushed aside or heated cooler gas along the line of the jets. The central temperature peak is clearly shown by both the radial and spectral mapping analysis, with the profile continuing to rise down to sub-kiloparsec scales. Despite this, the central entropy is very low and the central cooling time short, suggesting that the IGrM may be cooling rapidly within the galaxy. Support for cooling is provided by the filamentary \Ha +[N\textsc{ii}] nebula, which occupies the central few kiloparsecs of the galaxy, and whose longest filament extends along roughly the same axis as the radio emission, toward a region of low temperatures and entropies northwest of the core. This alignment, and the low velocity of the ionized gas with respect to the galaxy, suggest that the filament consists of material that was either uplifted during a past episode of lobe inflation, or has condensed from thermally unstable IGrM drawn up or disturbed by the AGN outburst. However, we see no clear cavities, and the radio data do not show any clear lobes. In this section we will therefore try to constrain the state of the group, and examine whether it is consistent with relaxation after an AGN outburst.

\subsection{IGrM cooling and thermal instability}
\label{sec:instability}
Early idealized simulations predicted that cold gas can precipitate out of the ICM in regions where the ratio of the cooling time to the free-fall time (t$_{\rm cool}$/t$_{\rm ff}$) falls below a threshold close to 10, if it is subject to perturbations \citep{McCourtetal12,Sharmaetal12}. Subsequent work demonstrated that AGNs can provide the necessary perturbations in this thermally unstable gas \citep{BanerjeeSharma14,Prasadetal15,Lietal15,Prasadetal17} either directly while their jets are actively interacting with the IGrM, or via the turbulence and gas motions induced by the buoyant rise of radio lobes, which can continue after the jets have ceased. In a turbulent ICM (or IGrM), condensation of cool gas is expected if the ratio of the cooling time to the eddy turnover time (t$_{\rm cool}$/t$_{\rm ff}$) is close to 1 \citep{Gasparietal18}. Observations of AGN in groups and clusters have generally confirmed the expectations from simulations with nebular emission and star formation typically found in systems with mimum values of t$_{\rm cool}$/t$_{\rm ff}$$\lesssim$20 and/or t$_{\rm cool}$/t$_{\rm ff}$$\sim$1, and closely associated with AGN-driven structures \citep[e.g.,][among many others]{Voitetal15,McNamaraetal16,Hoganetal17,Olivaresetal19,Olivaresetal22}.
Other sources of perturbation should also trigger the precipitation from thermally unstable gas, with simulations showing that condensation is also expected in the wakes of galaxies orbiting within the cluster or group \citep{Choudhuryetal19,Saeedzadehetal23}.

The presence of the \Ha +[N\textsc{ii}] filamentary nebula suggests that material is cooling from the IGrM in NGC~777. The velocity distribution of the ionized gas is inconsistent with acquisition via a gas-rich galaxy merger or stellar mass loss. The alignment of the filaments with the X-ray and radio structures suggests a connection with past AGN activity in the galaxy. We therefore calculate profiles of these ratios for NGC~777.

\begin{figure}
\includegraphics[width=\columnwidth,bb=35 210 560 760]{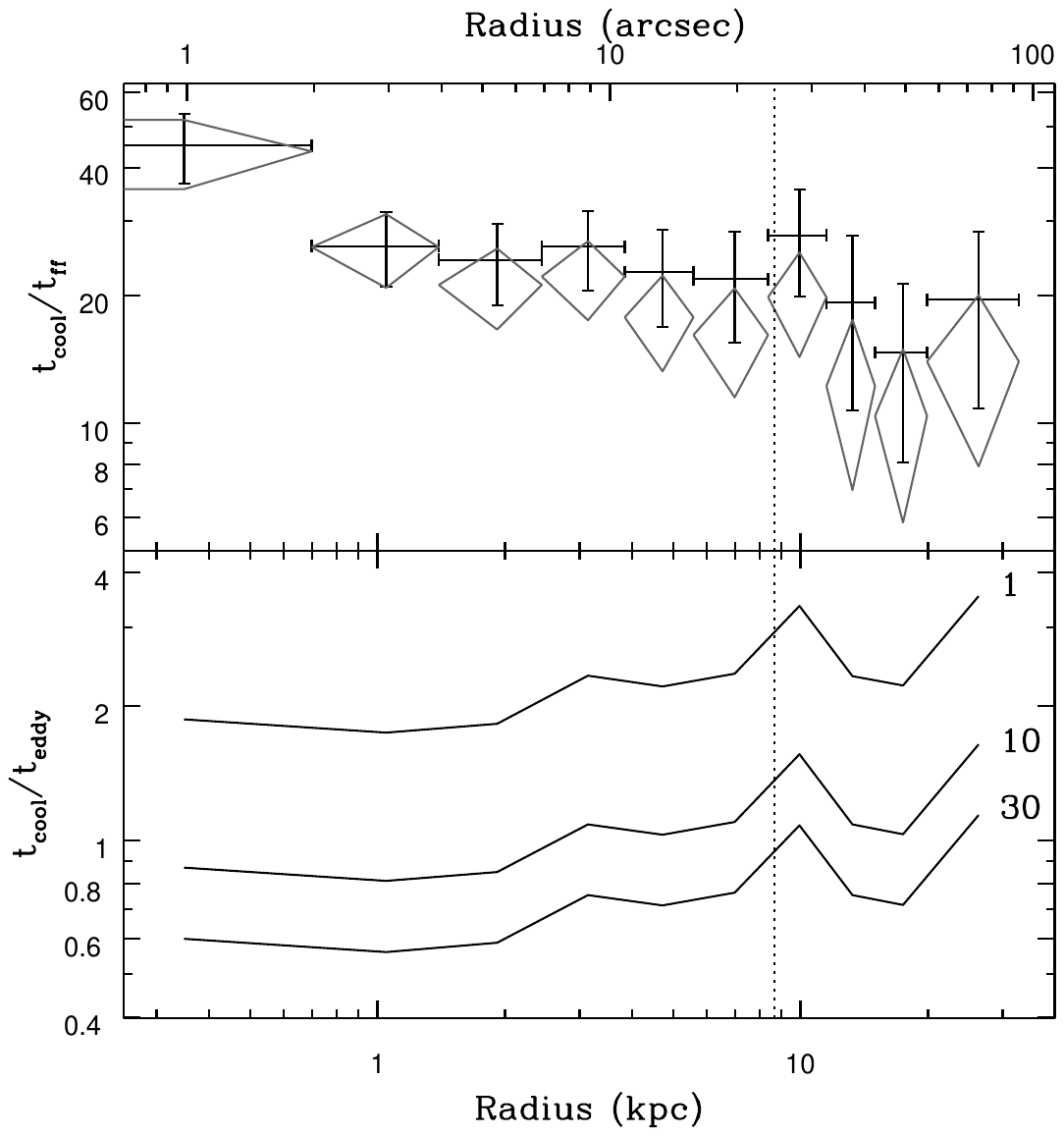}
\caption{\label{fig:tctff}Profiles of the ratio of isochoric cooling time to free-fall time (\textit{upper panel}) and eddy turnover time (\textit{lower panel}). The t$_{\rm cool}$/t$_{\rm ff}$ ratio is estimated from the \chandra\ data assuming hydrostatic equilibrium (black error bars) or from the stellar velocity dispersion profile of the galaxy (grey diamonds). The t$_{\rm cool}$/t$_{\rm eddy}$ profiles are labelled with the injection length scale $L$ (in kiloparsecs) used in their calculation. The vertical dotted line indicates the maximum extent of the \Ha +[N\textsc{ii}] emission.}
\end{figure}

We follow two approaches to estimate the free-fall time. \citet{Voitetal15} demonstrated that where the gravitational mass profile is dominated by a stellar component, as in giant elliptical galaxies, free-fall time is related to the stellar velocity dispersion $\sigma_*$ as t$_{\rm ff} \simeq r/\sigma_*$. \citet{Loubseretal18} found that the stellar velocity dispersion profile within the central $\sim$10~kpc of  NGC~777 is reasonably well described by a power law, $\sigma(r)=\sigma_o r^{\eta}$, where $r$ is the radius in kiloparsecs, the central velocity dispersion $\sigma_o$=321$\pm$9\kmps, and $\eta$=-0.078$\pm$0.01. We extrapolate the profile out to $\sim$30~kpc, to match the \chandra\ t$_{\rm cool}$ profile, but caution that the assumption that stars dominate the mass budget may not be accurate over the full range. Alternatively, under the assumption of a spherical halo in hydrostatic equilibrium, we can estimate the gravitational acceleration $g$ from the IGrM pressure and gas density profiles \citep{Liuetal19}, since

\begin{equation}
g = \frac{d\Phi}{dr} = \frac{1}{\rho}\frac{dP}{dr}
\end{equation}

\noindent where $\Phi$ is the gravitational potential, P is the IGrM pressure, and $\rho$ is the gas density. We describe the pressure profile with a generalized NFW model to ensure that the pressure gradient is well-behaved. The free-fall time is $\sqrt{2r/g}$.

Figure~\ref{fig:tctff} shows the isochoric t$_{\rm cool}$/t$_{\rm ff}$ profiles calculated by both methods, which are in good agreement within the uncertainties. The profiles both peak in the core with values of t$_{\rm cool}$/t$_{\rm ff}$ $\sim$40, show a shallow declining gradient in the 1-10~kpc radius range with values 17-25, and then fall to their lowest values at 10-20~kpc. The minimum value based on velocity dispersion is min(t$_{\rm cool}$/t$_{\rm ff}$)=10.4$\pm$4.6, while assuming hydrostatic equilibrium gives min(t$_{\rm cool}$/t$_{\rm ff}$)=14.7$\pm$6.6, both in the 15-19.9~kpc bin. Values below 20 have been associated with evidence of cooling at other wavelengths \citep[e.g.,][]{VoitDonahue15,Voitetal15}. This suggests that the gas is most likely to be most thermally unstable at 10-20~kpc radius, on the same scale as the ``butterfly wing'' surface brightness structures in the IGrM, and most thermally stable in the central kiloparsec.

In the chaotic cold accretion model of \citet{Gasparietal18}, condensation is expected to occur where t$_{\rm cool}$/t$_{\rm eddy}$ approaches unity. The eddy turnover timescale depends on the injection length scale $L$ and the velocity dispersion of the turbulence at the injection scale, $\sigma_{v,L}$. Neither parameter can be  measured directly for NGC~777 from current data, but $\sigma_{v,L}$ is expected to be similar to the velocity dispersion of any cooling gas seen at other wavelengths, such as the \Ha +[N\textsc{ii}] nebula. The central line of sight velocity dispersion of the \Ha\ in NGC~777 is 190$\pm$2\kmps\ \citep{Olivaresetal22}, falling to 77$\pm$3\kmps\ in the filaments, so including a factor of $\sqrt{3}$ to convert from line of sight to three-dimensional velocity dispersion, we adopt $\sigma_{v,L}$=133$\pm$5\kmps. The injection length scale $L$ may be similar to the size of the filaments or radio emission (both extending $\sim$10~kpc from the nucleus) or possibly to the scale of the surface brightness structures in the IGrM ($\sim$20~kpc). However, since the scale is unclear we have chosen to estimate t$_{\rm cool}$/t$_{\rm eddy}$ profiles assuming $L$=1, 10, or 30~kpc. Figure~\ref{fig:tctff} shows these profiles. The ratio may be close to or less than unity over much of the range of the profile, suggesting that the IGrM may be thermally unstable throughout the core of the NGC~777. Unlike t$_{\rm cool}$/t$_{\rm ff}$, these profiles suggest the gas is likely to be most thermally unstable in the central few kiloparsecs.

\subsection{The centrally peaked temperature profile}

\begin{figure}
\includegraphics[width=\columnwidth,bb=35 210 560 740]{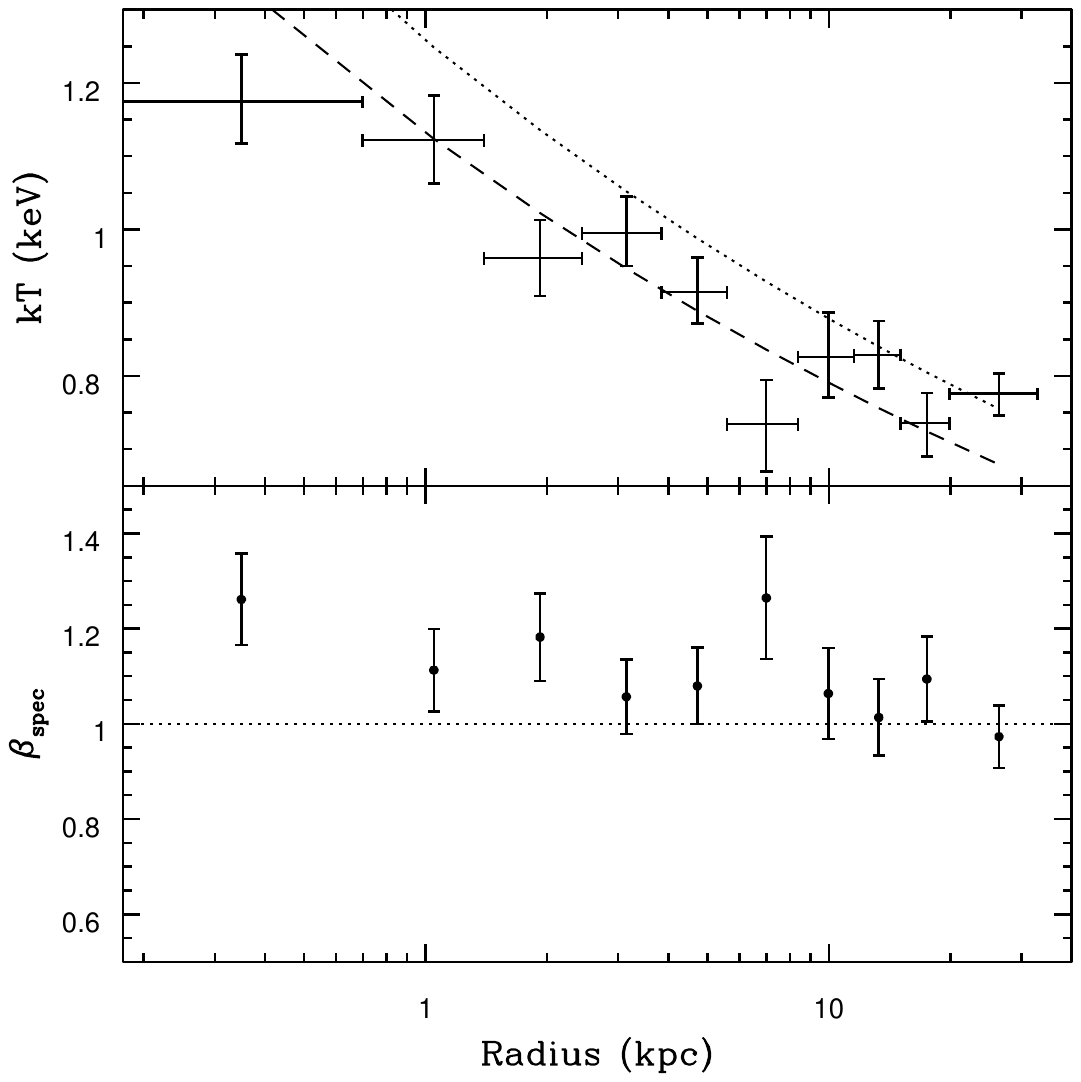}
\caption{\label{fig:Bspec}Comparison of IGrM temperature and stellar velocity dispersion in NGC~777. The \textit{upper panel} shows the \chandra\ deprojected temperature profile compared with the extrapolated velocity dispersion profile (dotted line) or the profile scaled by a factor of 0.9 (dashed profile). The \textit{lower panel} shows the parameter $\beta_{spec}$, the ratio of the kinetic energy of the stellar population to the thermal energy of the IGrM.}
\end{figure}

Figure~\ref{fig:Bspec} shows a comparison of the specific energies of the IGrM and stellar population of NGC~777 in the central $\sim$30~kpc of the galaxy. The ratio of these two energies, $\beta_{spec}$, is defined as

\begin{equation}
\beta_{spec} = \frac{\mu m_p \sigma_r^2}{kT} ,
\end{equation}

\noindent where $\mu$ is the mean molecular weight, $m_p$ is the proton mass, $\sigma_r$ is the line of sight velocity dispersion of the stellar population, and $kT$ is the temperature of the IGrM \citep[e.g.,][]{Sarazin88}. While this ratio was historically calculated for systems as a whole \citep[e.g.,][]{NaginoMatsushita09} we use the deprojected temperature profile and stellar velocity dispersion profile \citep[described in Section~\ref{sec:instability},][]{Loubseretal18} to examine its variation with radius. Figure~\ref{fig:Bspec} shows that for NGC~777 $\beta_{spec}$ is $\sim$1.1 over the radial range considered, indicating that the kinetic energy of the stellar population is $\sim$10\% greater than the thermal energy in the IGrM. Another way to consider this is to determine the temperature profile equivalent to the velocity dispersion profile assuming $\beta_{spec}$=1. Figure~\ref{fig:Bspec} shows that the shape of the profile is a good match for the deprojected temperature profile (except perhaps in the central bin) and that it only requires scaling down by $\sim$10 per cent (equivalent to $\beta_{spec}$=1.11 to match the normalization. This suggests that the IGrM gas and stellar population are in (or close to) equilibrium.

\subsection{AGN heating}
\label{sec:AGN}
While the radio and X-ray images show neither radio lobes nor clear cavities, the structures we observe are suggestive of past jet activity by the AGN of NGC~777. The extension of the radio emission to the NW and SE seems likely to be the fading remnant of old jets, while the butterfly-shaped X-ray morphology could be explained if radio lobes were inflated by these jets, pushing aside relatively dense, cool IGrM gas which now forms the butterfly's wings.

\begin{figure}
\includegraphics[width=\columnwidth,bb=55 150 560 650]{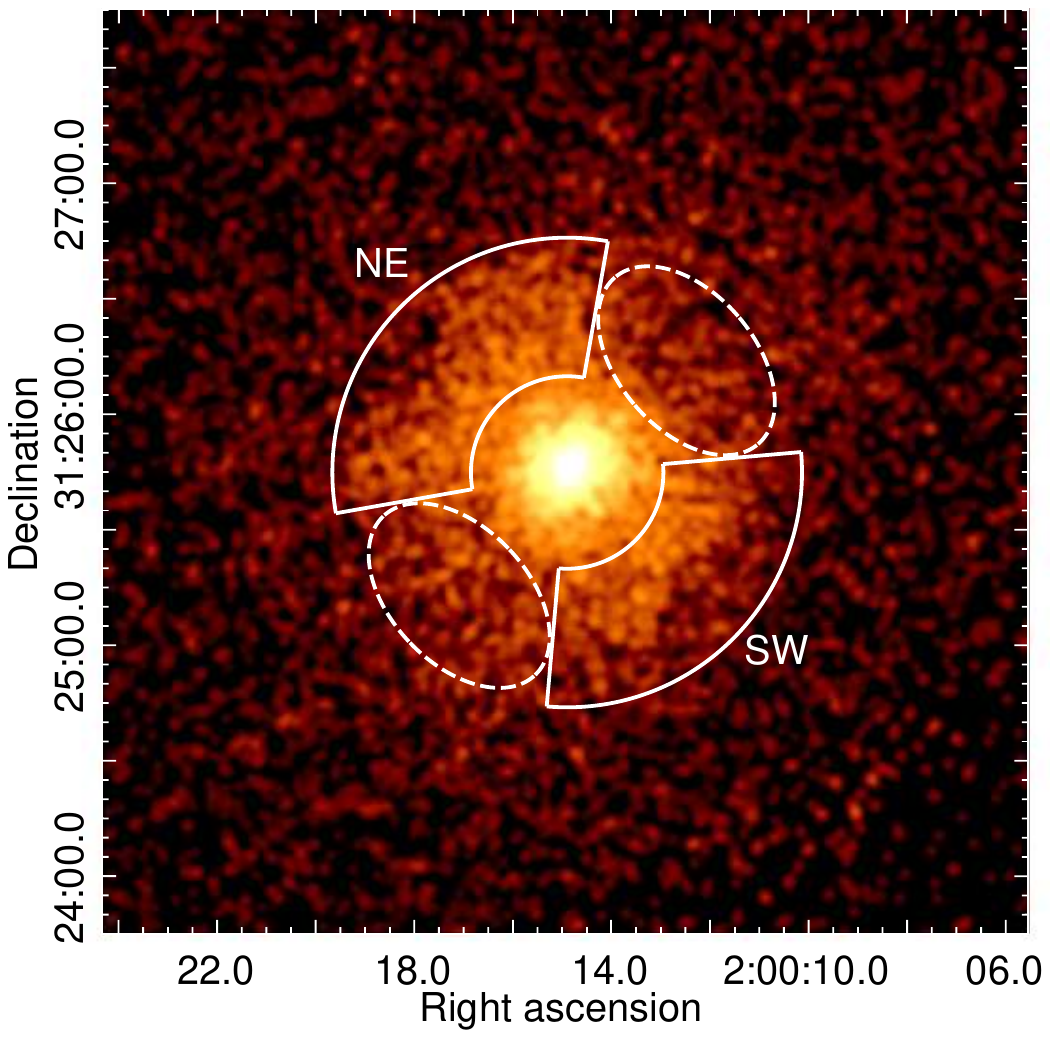}
\caption{\label{fig:cavs}\chandra\ 0.5-2~keV image with point sources removed and refilled, smoothed with a 2.5\arcs\ FWHM Gaussian. The dashed ellipses mark the regions chose to test the heating power of cavities associated with a past episode of AGN heating, but note that there is no firm detection of cavities in the system. The partial annuli labelled NE and SW indicate the regions used for comparison. For scale, the inner and outer radii of the partial annuli are 24.6\arcs\ (8.7~kpc) and 60\arcs\ (21.2~kpc).}
\end{figure}

Figure~\ref{fig:cavs} shows the \chandra\ 0.5-2~keV image of the group core, after removal and refilling of point sources. We do not see the clear surface brightness deficits or bright rims of compressed gas necessary to confirm the presence of cavities, and cavities beyond $\sim$1\arcm\ ($\sim$21~kpc) from the nucleus would likely not be detectable, owing to the relatively low surface brightness beyond that radius in the \chandra\ data. If we assume that the butterfly-shaped structure is the result of cavity inflation on 10-20~kpc physical scales, then the approximate size of such cavities is indicated by the dashed ellipses.

While cavities are not detected, we can test whether cavities of the size we have assumed are consistent with the data. Using the deprojected temperature and density at the radius of the potential cavities, we can use the APEC thermal plasma model to estimate the expected deficit in detected counts if their volume contains no IGrM plasma. We find that we would expect a deficit of $\sim$290 counts in each cavity region, in the 0.5-2~keV band. Measuring the actual surface brightness in those regions, we compare it with the surface brightness in the NE and SW regions. We find that the cavity regions contain 669 (NW) and 683 (SE) counts, whereas scaling from the average surface brightness in the partial annuli we would expect to see 946 and 944 counts respectively. These differences in measured surface brightness in and outside the possible cavity regions are $\sim$8$\sigma$ significant. The observed deficits (277 and 261 counts respectively) are somewhat smaller than the $\sim$290 counts expected if cavities are present, but they are consistent with that prediction within 1$\sigma$ uncertainties. Given the uncertainties involved in the calculation (cavity volume and filling factor, variations in surface brightness between and within the NE and SW comparison regions, variation and uncertainties on spectral properties across the radial range of the cavities) we can only say that the imaging is consistent with the presence of cavities on this scale, though the lack of clear rims means that we cannot be sure such cavities exist.

We note that smaller cavities or channels closer to the core of the galaxy may be present, but the higher surface brightness in this area means that large cavities should be easily detectable. For example, $\sim$4~kpc ($\sim$11.5\arcs) radius cavities 5~kpc from the nucleus should be detected at $\sim$13$\sigma$ significance. Any inner cavities or channels are thus likely to be small and contribute little to any heating.

To determine the energy required to create cavities on the scale we have assumed, we extracted spectra from concentric annuli, excluding regions within the angular ranges covered by the potential cavities (5-80\degree\ and 190-265\degree, defined anti-clockwise from due west) at radii larger than 25\arcs\ (8.9~kpc). The angular ranges used outside the galaxy core are indicated in Figure~\ref{fig:cavs} by the regions labelled SW and NE. The width of the annuli was chosen to ensure a S/N$\geq$25 in the spectra, to ensure a similar radial resolution to our azimuthally averaged profiles for the group as a whole (see Section~\ref{sec:profiles}). We then repeated our deprojection analysis to measure the IGrM pressure in the brighter wings of the butterfly structure. This value was then adopted as the pressure against which any jets would have to have worked to inflate cavities. Table~\ref{tab:cavs} gives the physical scales of the potential cavities, the surrounding pressure $p$ and the implied enthalpy, under the usual definition of cavity enthalpy as $4pV$ where $V$ is the cavity volume. We approximate the cavities as oblate ellipsoids viewed from the side, so that their depth along the line of sight is equal to their width normal to the axis from the cavity center to the AGN. Uncertainties on enthalpy are estimated as described in \citet{OSullivanetal11b}. As the potential cavity regions have the same size and are at the same radius, their enthalpies are identical.

\begin{table*}
\caption{\label{tab:cavs} Parameters used in estimating the enthalpy of the potential cavities in NGC~777, where R is the radius of the midpoint of the cavities from the AGN, r$_{\rm maj}$ and r$_{\rm min}$ are the semi-major and semi-minor axes of the ellipses approximating the cavity regions. As the two potential cavity regions have identical radii, values for a single cavity are shown.}
\begin{center}
\begin{tabular}{cccccc}
\hline
\hline
R & r$_{\rm maj}$ & r$_{\rm min}$ & Volume & Pressure & Enthalpy \\
(arcs) & (arcs) & (arcs) & (cm$^3$) & (erg cm$^{-3}$) & (erg) \\
\hline
44 & 28.5 & 17.7 & 7.84$\times$10$^{67}$ & 1.58$^{+0.23}_{-0.22}$$\times$10$^{-11}$ & 4.97$^{+3.13}_{-1.00}$$\times$10$^{57}$ \\[+1mm]
\hline
\end{tabular}
\end{center}
\end{table*}

The total enthalpy expected for a pair of cavities on this scale is $\sim$9.9$\times$10$^{57}$~erg, sufficient to balance radiative cooling in the central $\sim$20~kpc (1.31$\times$10$^{42}$\ergps) for 240~Myr. This assumes that any cavities were filled with lobes of relativistic plasma from AGN jets, and that all of the energy in the particles and magnetic fields within those lobes is available to heat the IGrM. 


\subsubsection{Timescales of AGN heating}
The timescale of formation of cavities at these locations is difficult to determine. Since clear cavities and lobes are not detected, any such structures must be old, with jet activity having ceased long enough ago that radio emission has had time to fade below detectability even at low frequencies, and any rims of compressed IGrM gas have diffused away. As the plasma of radio jets and lobes radiates away its energy most efficiently at high frequencies, aging is expected to produce a break in its spectrum, with steeper spectral indices above a break frequency $\nu_{\rm br}$. The radiative age, $t_{\rm rad}$, of the plasma (i.e., the time since particle acceleration ceased) is dependent on the break frequency and the equipartition magnetic field strength:

\begin{equation}
t_{\rm rad} = 1590\frac{B_{\rm eq}^{0.5}}{B_{\rm eq}^{0.5}+B_{\rm CMB}^2}[(1+z)\nu_{\rm br}]^{-0.5}
\end{equation}

\noindent where $B_{\rm CMB}$=3.2(1+$z$)$^2$ is the equivalent magnetic field of the cosmic microwave background at the redshift of the source, $z$, with magnetic field strengths in units of $\mu$G, frequencies in GHz and timescales in Myr. Since we observe no break in the spectrum, we adopt 140~MHz as an upper limit on the break frequency. We do not know the magnetic field strength in NGC~777, but values of order 5~$\mu$G are typical for large (tens of kiloparsec) scale lobes in groups and clusters. We find that for $B_{\rm eq}$=5~$\mu$G (1$\mu$G, 10$\mu$G), the lower limit on the age is $t_{\rm rad}$$\geq$262~Myr (120~Myr, 353~Myr). However, this estimate ignores a number of factors which could affect spectral aging, most notably adiabatic expansion of the radio plasma, which can cause the break frequency to either increase or decrease, depending on whether the expansion takes place while new particles are still being injected, or afterwards \citep[e.g.,][]{ScheuerWilliams68,BlundellAlexander94,Murgiaetal99}. Since we have no way to constrain the history of such expansion in the source, we cannot correct for it.

We can also estimate characteristic dynamical timescales for cavities such as the sonic timescale and refill time \citep[see, e.g.,][]{Birzanetal04}. Defining the sonic timescale as the time taken for the outer edge of a cavity to expand from the nucleus to the assumed maximum radius ($\sim$21~kpc) at the sound speed of the IGrM, we adopt 0.8~keV as a representative IGrM temperature and find a timescale of 44~Myr (for a sound speed of 464\kmps). The refill timescale for cavities of this size is slightly larger, 57~Myr. Both of these timescales are short enough that, if they represented the actual time since the AGN outburst, we would likely still be able to detect emission from radio lobes and for expansion at the sound speed we would certainly expect to see evidence of IGrM compression. Lobe emission might fade more rapidly than expected if the magnetic field strength were unusually high, causing rapid aging, or if a significant fraction of the relativistic particles were able to diffuse out of the lobes into the IGrM, but we would still expect to see cavity rims. The true timescale of any outburst must be significantly longer.

It therefore seems likely that any past outburst occurred at minimum $\sim$50~Myr ago, and probably $\sim$200~Myr ago. An outburst on the scale we have assumed would provide sufficient energy to balance cooling on this timescale, assuming that the enthalpy of the outburst is efficiently coupled to the IGrM. However, it is less clear how such an outburst would produce the centrally peaked temperature profile we observe. Shock heating is a possibility, but in groups where AGN-driven shocks are observed, visible rises in temperature appear to be short-lived compared to the timescales we are considering, and visible only close to the shock front itself. Longer lasting mechanisms, such as turbulence driven by the motions associated with the rise and expansion of cavities, or heating by relativistic particles leaking from old radio structures, might be more suitable.

\subsection{Supernova heating}
\label{sec:snae}
Another source of energy to heat the IGrM within the galaxy are supernovae (SNae). The total bolometric X-ray luminosity of the IGrM within $\sim$90\arcs\ ($\sim$32~kpc) radius is 1.72$^{+0.16}_{-0.15}$$\times$10$^{42}$\ergps. If we conservatively define the hot core region to have a radius $\sim$16\arcs\ ($\sim$5.6~kpc) the bolometric luminosity of the core is 5.14$\pm$0.33$\times$10$^{41}$\ergps. For the dominant 5~Gyr old stellar population, we would expect a SNIa rate of $\sim$0.007~yr$^{-1}$ \citep{Rodneyetal14}, adopting a $K$-band mass-to-light ratio of unity \citep{LonghettiSaracco09} and luminosity log L$_K$=11.34 \citep{EllisOSullivan06}. Assuming each SN releases 10$^{51}$~erg of energy \citep{Bethe94} this is equivalent to an energy injection rate of 2.22$\times$10$^{41}$\ergps. Additionally, for a young population arising from the estimated $\sim$0.2\Msolpyr\ of star formation in NGC~777, we can expect roughly one SNII for every 100\Msol\ of stars formed, with energy injection rate 6.34$\times$10$^{40}$\ergps. Combining these two classes of SNae, the total expected energy injection rate 2.85$\times$10$^{41}$\ergps. This is significantly less than the X-ray luminosity of the core, and the discrepancy will be exacerbated once the (likely $\lesssim$10\%) efficiency with which SNae heat the IGrM is taken into account \citep[][]{KravtsovYepes00}. It is therefore clear that SN heating cannot be the cause of the hot core in NGC~777.

\subsection{Stellar wind heating}
\label{sec:thermalize}
In massive early-type galaxies like NGC~777 gas lost from stars through stellar winds is a potentially significant source of heating. The gas in these winds shares the velocity of the parent star, and when it interacts with the surrounding hot medium its kinetic energy may be thermalized \citep{Mathews90}. This process has been explored by \citet{Conroyetal15} who find that the gas is expected to thermalize at a temperature kT$\sim$$\sigma_*^2$, where $\sigma_*$ is the 3-dimensional velocity dispersion of the stellar population. This could provide a natural explanation of the apparent correlation between the specific energies of the hot gas and stellar population in NGC~777.

We calculate the maximum heating rate available, using \citet{Conroyetal15} Eqn.~3, and find that for the galaxy as a whole, we would expect a maximum rate of 2.9$\times$10$^{41}$\ergps. This is $\sim$17\% of the energy input required to balance cooling in the central $\sim$30~kpc. However, the maximum heating rate will only be achieved when kT$\ll$$\sigma_*^2$, i.e., when the kinetic energy of the stellar winds greatly exceeds the temperature of the surrounding medium. When the IGrM temperature exceeds $\sigma_*^2$, there will be a net cooling effect, as energy is required to heat the stellar wind material to the temperature of the surrounding medium. \citet{Conroyetal15} suggest that this will tend to result in systems with core temperatures similar to $\sigma_*^2$, as hotter systems are cooled by this process, and those which cool below $\sigma_*^2$ are heated. In NGC~777, as the IGrM temperature follows the profile of $\sigma_*^2$ in the core, but falls $\sim$10\% below it, the current heating rate from this mechanism will be only a fraction of the maximum rate we have calculated. This suggests that the heating effect of stellar wind thermalization is likely to be minimal at present, and even at its maximum rate would not be able to balance cooling.

\subsection{Gravitational heating}
As noted by \citet{Lewisetal00}, as the IGrM radiates away energy and contracts within the gravitational potential of the group, it will be subject to PdV work. Where this exceeds the radiative losses, the temperature of the gas can rise as it moves inward to the group core, and for a sufficiently concentrated mass profile and rapid inflow, this will produce a central temperature peak. Large-scale simulations naturally incorporate gravitational heating in evolving systems alongside star formation and AGN feedback. For X-ray luminous galaxy groups, we expect periodic feedback from the central AGN, which will drive gas out of the group core, reducing IGrM density and cooling rate. In the periods between such outbursts, the IGrM will radiate away the injected energy and contract \citep[see, e.g.,][]{Cieloetal18} and during this phase gravitational heating will play a role.

\citet{Jungetal22} present temperature profiles from a sample of systems drawn from the \textsc{Romulus} suite of simulations. These are typically less massive than NGC~777 (M$_{200}$$\sim$10$^{12}$-10$^{13}$\Msol ) but it is notable that for systems in which star formation has been quenched in the dominant galaxy, the temperature profiles almost universally show a broad central peak. In NGC~777 we see a factor of $\sim$1.5 increase in temperature from 0.1R$_{500}$ (43~kpc or 123\arcs) into the core; this is consistent with the temperature gradients seen in some of Jung et al.'s systems. It should be noted that while the \textsc{Romulus} simulation was calibrated to ensure a realistic cosmic star formation history, not the physics of the IGrM, the scaled entropy profiles of the systems are consistent with those of (somewhat more massive) observed groups. However, in higher mass halos the simulation tends to overpredict star formation. This may indicate flaws in the representation of the cycle of IGrM cooling and AGN feedback.

\citet{Khosroshahietal04} used a simple steady-state model to demonstrate that gravitational heating can reproduce the centrally-peaked temperature profile of NGC~6482. In such a model, the temperature profile of gas within some outer radius will depend on the starting temperature at that radius, the change in gravitational potential between that radius and the radius of interest, and the radiative losses over the time taken for the gas to flow from the outer radius to the inner. \citet{OSullivanetal17} further tested this model against the temperature profiles of several high-richness groups from the Complete Local-volume Group Sample (CLoGS), showing that it could approximate some of those with central temperature peaks, but that the cooling inflow rates were large compared to the typical masses of cold gas observed in such systems.

Comparing this model to the temperature profile of NGC~777, we find that it does not reproduce the shape of the profile. At inflow rates of a few \Msolpyr\ radiative cooling dominates, resulting in a central temperature decline. At higher rates, heating begins to dominate in the central few kiloparsecs, but the temperature profile is too sharply peaked in the core. A high inflow rate of $\sim$15\Msolpyr\ is required to produce a peak of the required magnitude, but does not reproduce the width of the observed temperature peak. If we adopt the cooling rate found in the HCF model fit to the RGS data, $\dot M$$\simeq$38\Msolpyr, the maximum temperature of the central peak is roughly double the observed central temperature and the peak is too broad.

This steady state inflow model is over-simplistic as a representation of the period of relaxation between AGN outbursts. Since outbursts are expected to reduce IGrM density in the inner part of the group, we would expect somewhat lower radiative cooling rates immediately after an outburst, increasing with time as gas is able to flow back into the core. The model also does not include the effects of other heating sources, though as shown in Section~\ref{sec:snae} and \ref{sec:thermalize}, heating from supernovae and stellar winds appears to provide only a minimal contribution in NGC~777 at present. However, the inability of the model to reproduce the observed shape of the temperature profile at any inflow is worrying, and the need for relatively high rates to produce a central temperature comparable to that observed is also problematic. If, as seems likely, the time since the last AGN outburst is $>$10$^8$~yr, we would expect a significant build-up of cooled gas in the galaxy core. While gravitational heating may explain the centrally-peaked profiles of some groups, it is unclear whether it can in NGC~777.

\section{Summary and Conclusions}

Whereas cool-core galaxy clusters typically show both short central cooling times and temperature profiles which decline in to small radii, some galaxy groups and individual ellipticals with short central cooling times show centrally-peaked temperature profiles. NGC~777 is a prime example, with a central cooling time of $<$100~Myr, but a temperature profile that rises from $\sim$0.8~keV at 10-20~kpc radius to $\sim$1.2~keV in its core. VLT MUSE observations provided evidence of cooling in the form of a $\sim$10~kpc filamentary \Ha +[N\textsc{ii}] nebula in the galaxy core, but only ambiguous indications of feedback were seen, with no confirmed detections of cavities or radio lobes. To investigate the origins of the hot core, we acquired new \chandra\ ACIS X-ray observations of the group-dominant elliptical, supported by a deep low-frequency radio observation from the uGMRT to search for any evidence of past AGN outbursts. We summarize our findings below:

\begin{enumerate}
\item A radial spectral deprojection confirms the previous finding that NGC~777 has a centrally peaked temperature profile, with no sign of any central decrease down to scales $\sim$700~pc. Despite this, the entropy and cooling time in the galaxy core are strongly suggestive of cooling, with a central entropy of 2.62$^{+0.19}_{-0.18}$\kevcmsq\ and central cooling time 71.3$^{+12.8}_{-13.1}$~Myr. The entropy at 10~kpc is K$_{10}$=32.61$^{+4.37}_{-5.01}$\kevcmsq.

\item X-ray imaging reveals structure in the hot gas component within the galaxy. A bright central core is bordered by ``butterfly-wing'' high surface brightness structures to the northeast and southwest (on the minor axis of the elliptical) with fainter regions to the northwest and southeast (on its major axis). Spectral mapping suggests that the ``wings'' of brighter emission are associated with cooler temperatures, while channels of higher temperatures extend from the central temperature peak along the galaxy's major axis. While entropies are relatively low throughout much of the galaxy, the very lowest entropies are found in the galaxy core and to its immediate northwest, along the line of the longest filament of the \Ha +[N\textsc{ii}] nebula. This correlation suggests that these lowest entropy regions are the locus of cooling and condensation in the IGrM, from which the ionized gas has formed.

\item Our uGMRT band~3 (300-500~MHz) observation shows low-level diffuse radio emission surrounding the AGN, extending to $\sim$28\arcs\ radius ($\sim$10~kpc) along its longest axis and aligned northwest-southeast. This confirms the finding of diffuse emission on this axis from LOFAR at 140~MHz, and is similar to the alignment and (projected) length of the longest \Ha +[N\textsc{ii}] filament.

\item Using archival \xmms\ RGS spectra, we tested whether NGC~777 could harbor a hidden (i.e., partially intrinsically absorbed) cooling flow (HCF). The HCF model and a simpler 2-temperature plasma model fit the RGS spectra reasonably well, but the covering fraction of the intrinsic absorption component of the HCF model is unconstrained. When using minimally binned spectra, both models suggest stronger smoothing of the emission from the cooler component, implying that the hotter gas is more centrally concentrated, consistent with the temperature distribution we observe in the \chandra\ data. We conclude that the RGS spectra alone are not capable of discriminating between the temperature profile we observe in the \chandra\ data and a hidden cooling flow.

\item We examine the thermal stability of the IGrM and show that the ratio t$_{\rm cool}$/t$_{\rm ff}$ peaks in the galaxy core, falling at larger radii with minimum values of 10-15 at $\sim$15-20~kpc ($\sim$42-56\arcs) radius. This is a little larger than the extent of the \Ha +[N\textsc{ii}] filaments, but suggests that gas is more likely to be precipitating from thermal instabilities at the outer edge of the nebula than in the core. By contrast, profiles of t$_{\rm cool}$/t$_{\rm eddy}$ show their lowest values in the core, though the profiles are relatively flat over most of the range out to 20~kpc radius.

\item The slope of the temperature profile within $\sim$10~kpc is similar to that of the kinetic energy profile of the stellar population (based on the velocity dispersion profile) with only $\sim$10\% difference in normalization. The stellar velocity dispersion is a good tracer of mass within the galaxy, so this suggests that the hot gas in this region is also in equilibrium with the gravitational potential. Given the short cooling times in the galaxy core, this is surprising. Defining the central temperature peak as extending to $\sim$20~kpc, the energy input required to balance radiative cooling is 1.31$\pm$0.11$\times$10$^{42}$\ergps.


\item Examining possible sources of heating, we find while thermalization of stellar winds would, in the absence of cooling, produce a temperature profile correlated with the stellar kinetic energy profile, the energy available from this source and from supernovae is much too small to balance radiative cooling. Simulations have shown that gravitational heating associated with the contraction and inflow of gas after the last AGN outburst $\sim$200~Myr ago has the potential to produce centrally peaked temperature profiles with properties similar to those we observe in NGC~777. However, a simple steady-state model suggests that an implausibly large inflow rate ($\sim$15\Msolpyr) would be required to produce a peak of the magnitude we observe, and even at this rate the model fails to reproduce the breadth of observed profile, producing only a narrow temperature peak in the core. While we do not observe radio lobes or detect clear cavities, we find that the X-ray surface brightness distribution is consistent with the presence of cavities $\sim$16~kpc northwest and southeast of the core, with the brighter ``butterfly wing'' regions formed from denser gas pushed aside during cavity inflation. The enthalpy of such cavities would be $\sim$9.9$\times$10$^{57}$~erg, sufficient balance cooling in the central $\sim$20~kpc for 240~Myr. If cavities on this scale exist, they must be in equilibrium with their surroundings, old enough that any compressed gas rims have dissipated, and their radio lobes have faded beyond detectability even in the 140~MHz LOFAR data. If they exist, observations at frequencies $<$100~MHz would likely be needed to detect them. Cavities on larger scales are also possible, with the low surface brightness regions perhaps only showing us their inner boundary. Such cavities would not be detectable in the available X-ray data.

\end{enumerate}

These results suggest that we may be observing NGC~777 at a particular phase of the cooling-feedback cycle, long enough after the AGN has shutdown that lobe emission and cavity rims have faded, but before radiative cooling erases the central temperature peak and a large reservoir of cooled gas built up. If so, this may imply that AGN heating was most effective in the central few kiloparsecs where the break in the entropy profile is observed, and that the hot core is the product of shock or particle heating. However, the origin of the similarity between the IGrM temperature profile and the stellar kinetic energy profile remains a mystery.

\begin{acknowledgments}
The authors thank the anonymous referee for suggesting the inclusion of the RGS data, and the staff of the \chandra\ X-ray Observatory and the uGMRT for their help with the observations used in this work. Support for this work was provided by the National Aeronautics and Space Administration through Chandra Award Number GO2-23118X issued by the Chandra X-ray Center, which is operated by the Smithsonian Astrophysical Observatory for and on behalf of the National Aeronautics Space Administration under contract NAS8-03060. The GMRT is run by the National Centre for Radio Astrophysics (NCRA) of the Tata Institute of Fundamental Research (TIFR).
\end{acknowledgments}

\facilities{CXO, GMRT, XMM}

\software{CIAO \citep[v4.15][]{Fruscioneetal06}, Sherpa \citep[v4.15][]{Freemanetal01}, XSpec \citep[12.13.0b][]{Arnaud96}, SPAM \citep[][]{Intemaetal09}, SAS \citep[][]{Gabrieletal04}}

\bibliographystyle{aasjournal}
\bibliography{../paper}

\end{document}